\documentclass[a4paper,12pt,numbers,sort&compress]{article}

\usepackage{amsmath,amssymb,amsfonts,epsfig,cite,setspace}
\pdfoutput=1

\begin{document}


\vskip 0.25in

\newcommand{\todo}[1]{{\bf ?????!!!! #1 ?????!!!!}\marginpar{$\Longleftarrow$}}
\newcommand{\fref}[1]{Figure~\ref{#1}}
\newcommand{\nn}{\nonumber}
\newcommand{\tr}{\mathop{\rm Tr}}
\newcommand{\firr}[1]{{}^{{\rm Irr}}\!{\cal F}^{\flat}_{#1}}

\newcommand{\comment}[1]{}

\newcommand{\cM}{{\cal M}}
\newcommand{\cW}{{\cal W}}
\newcommand{\cN}{{\cal N}}
\newcommand{\cZ}{{\cal Z}}
\newcommand{\cO}{{\cal O}}
\newcommand{\cB}{{\cal B}}
\newcommand{\cC}{{\cal C}}
\newcommand{\cD}{{\cal D}}
\newcommand{\cE}{{\cal E}}
\newcommand{\cF}{{\cal F}}
\newcommand{\cX}{{\cal X}}
\newcommand{\IA}{\mathbb{A}}
\newcommand{\IP}{\mathbb{P}}
\newcommand{\IQ}{\mathbb{Q}}
\newcommand{\IR}{\mathbb{R}}
\newcommand{\IC}{\mathbb{C}}
\newcommand{\IF}{\mathbb{F}}
\newcommand{\IV}{\mathbb{V}}
\newcommand{\II}{\mathbb{I}}
\newcommand{\IZ}{\mathbb{Z}}
\newcommand{\re}{{\rm Re}}
\newcommand{\im}{{\rm Im}}
\newcommand{\li}{{\rm Li}}

\newcommand{\CA}{\mathbb A}
\newcommand{\CP}{\mathbb P}
\newcommand{\tmat}[1]{{\tiny \left(\begin{matrix} #1 \end{matrix}\right)}}

\newcommand{\diff}[2]{\frac{\partial #1}{\partial #2}}
\newcommand{\vev}[1]{\langle #1 \rangle}

\newcommand{\drawsquare}[2]{\hbox{%
\rule{#2pt}{#1pt}\hskip-#2pt
\rule{#1pt}{#2pt}\hskip-#1pt
\rule[#1pt]{#1pt}{#2pt}}\rule[#1pt]{#2pt}{#2pt}\hskip-#2pt
\rule{#2pt}{#1pt}}
\newcommand{\fund}{\raisebox{-.5pt}{\drawsquare{6.5}{0.4}}}
\newcommand{\antifund}{\overline{\fund}}

\newtheorem{theorem}{\bf THEOREM}
\def\thetheorem{\thesection.\arabic{theorem}}
\newtheorem{conjecture}{\bf CONJECTURE}
\def\thetheorem{\thesection.\arabic{conjecture}}
\newtheorem{observation}{\bf OBSERVATION}
\def\thetheorem{\thesection.\arabic{observation}}

\def\theequation{\thesection.\arabic{equation}}
\newcommand{\setall}{\setcounter{equation}{0}
        \setcounter{theorem}{0}}
\newcommand{\setequation}{\setcounter{equation}{0}}

\renewcommand{\thefootnote}{\fnsymbol{footnote}}
\centerline{\Huge On Fields over Fields}
~\\
\vskip 2mm
\centerline{
{\large Yang-Hui He}\footnote{\tt hey@maths.ox.ac.uk}
}
~\\
{\scriptsize
\begin{center}
\begin{tabular}{ll}
  $^1$ & {\it Rudolf Peierls Centre for Theoretical Physics, }
  {\it Oxford University, 1 Keble Road, OX1 3NP, U.K.}\\
  $^2$ & {\it 
       Merton College, Oxford, OX1 4JD, U.K.}\\
  $^3$ & {\it Department of Mathematics, City University,} 
  {\it Northampton Square, London EC1V 0HB, U.K.}\\
  $^4$ & {\it School of Physics, NanKai University,}
  {\it Tianjin, 300071, P.R.~China}
\end{tabular}
\end{center}
}
~\\
~\\

\begin{abstract}
We investigate certain arithmetic properties of field theories.
In particular, we study the vacuum structure of supersymmetric gauge theories as algebraic varieties over number fields of finite characteristic.
Parallel to the Plethystic Programme of counting the spectrum of operators from the syzygies of the complex geometry, we construct, based on the zeros of the vacuum moduli space over finite fields, the local and global Hasse-Weil zeta functions, as well as develop the associated Dirichlet expansions.
We find curious dualities wherein the geometrical properties and asymptotic behaviour of one gauge theory is governed by the number theoretic nature of another.
\end{abstract}

\newpage
\tableofcontents

\vskip 1in

\section{Prologus}\setall
On the number theoretical properties of the geometric structures arising from physics there has been growing interest.
Though perhaps relatively nascent a field in comparison to the tremendous cross-fertilization which algebraic and differential geometry have enjoyed with gauge and string theories, ever augmenting importance and ever increasing profundity of these number theoretic connections compel us.
Recent developments in investigating Calabi-Yau varieties over finite fields \cite{Candelas:2000fq,Candelas:2004sk}, modularity of zeta functions of Calabi-Yau manifolds \cite{modCY}, emergence of modular forms and Moonshine behaviour in mirror maps, Gromov-Witten invariants and matrix models \cite{K3,Aganagic:2006wq,He:2003pq}, and especially geometric perspectives on Langlands duality via S-duality \cite{Kapustin:2006pk} exemplify the incipience of arithmetic within a physical context, in particular with regard to the visions of the Langlands programme.

Our concern shall be supersymmetric gauge theory, the physics, the geometry, as well as the interplay thereof have established themselves as a canonical subject, of rich structure and fundamental importance.
The algebraic geometry of the theory has been crucial to such deep insight as the Holographic Principle. Indeed, in the AdS/CFT correspondence, wherein the gauge theory arises from the world-volume dynamics of branes, the space of vacua parameterizes precisely the bulk geometry, typically Calabi-Yau.
This vacuum moduli space (VMS) of gauge theories has itself been studied \cite{lt}, and with the advances in computational commutative algebra, been subject to new scrutiny \cite{Gray:2005sr,Berenstein:2002ge,Ferrari:2008rz}, uncovering perhaps unexpected signatures in such standard examples as the MSSM \cite{Gray:2006jb} or sQCD \cite{Gray:2008yu}.

A programme of enumerating operators in a supersymmetric gauge theory has recently been constructed, the methodology, dubbed the ``Plethystic Programme'', harnesses the algebraic geometry of the classical VMS, determined from certain flatness conditions once the Lagrangian is known \cite{Benvenuti:2006qr,Feng:2007ur}. Interpreting the vacuum as prescribed by an ideal in a, possibly complicated, polynomial ring, the Hilbert series and its plethystic exponentiation can then be regarded as grand canonical partition function which then encodes the chiral ring of operators of the gauge theory as a statistical-mechanical system.

We are therefore naturally endowed with two implements of which we feel obliged to take advantage: (1) a wealth of experimental data in the form of catalogues of supersymmetric gauge theories, each of which engendering an algebraic variety, realized as the VMS, and (2) an algorithmic framework within which the geometry of the VMS is exploited for the sake of enumeration of the spectrum of BPS operators and in which a fruitful dialogue between combinatorics and gauge theory is engaged.
This concurrent inspiration, experimental and theoretical, in further accordance with the aforementioned skein of number theory already weaving herself into the tapestry of algebraic geometry arising from physics, shall suffice to serve as a beacon to our path.

And the path on which we are led will be as follows.
We commence with a brief review in \S\ref{s:rev}, setting notation and laying the foundation, of the two chief protagonists.
First, we outline how to obtain the moduli space of vacua, given the ($\cN=1$ superspace) Lagrangian of an arbitrary supersymmetric gauge theory, by algorithmically recasting the F- and D-flatness conditions as a quotient of the space of (mesonic) gauge invariant operators (GIO) by the Jacobian ideal of the superpotential.
The VMS is then explicitly realized as an affine cone over a (weighted) projective variety.
The geometry, and in particular the Hilbert series of the VMS and combinatorial functionals thereof, is then utilized in the enumeration of the GIOs via the syzygies.
Second, of the vast subject of algebraic varieties over finite number fields we touch upon some rudiments which will be of use.
We emphasize on the construction of the local zeta function and the beautiful restriction of its rational form by the Weil Conjectures.
Then, forming the Euler product over primes, we remind the reader of the arrival at the global Hasse-Weil zeta function and its Dirichlet L-series expansion.

Thus armed, we march through a multitude of gauge theories in \S\ref{s:eg}, many of which are well-known, such as the free theory and SQCD. We will investigate these gauge theories under our new light, by finding the VMS, which is then subject to plethystic analyses, some of which have already been investigated through the Plethystic Programme in the literature, but more importantly, to the reduction over finite fields. A host of zeta functions is then constructed for these field theories, and various properties, observed.

Fortified by our gaining experience on the two enumerative problems, one counting the syzygies and the other, zeros over number fields, one could not resist but to perceive them, both deeply rooted in gauge theory, in a unified outlook.
We attempt, in \S\ref{s:compare}, to glimpse at this unifying principle, and show how one may proceed from one to the other, whereby establishing a curious duality between sets of gauge theories, with the geometric invariant properties of one controlling the arithmetic properties of another.
In the case where the VMS is dimension one and Calabi-Yau, {\it viz.}, the elliptic curve, this correspondence is very much in the spirit of the Modularity Theorem of Taniyama-Shimura-Weil-Wiles.
We trudge on, in \S\ref{s:asym}, towards the asymptotic behaviour of the operators in the gauge theory, already an integral component of the Plethystic Programme, but now accompanied by this duality, and see once more how the emergence of the Dirichlet series in governing the large R-charge and large $N$ trends in the physics. Finally, we part with concluding remarks and prospects in \S\ref{s:conc}.

\section{Dramatis Person\ae}\label{s:rev}\setall
Let us begin with a rather pedagogical presentation of the two subjects which will be crucial to our ensuing investigations. The contrast and parallels between them will constitute the comparative study in which we shall engage.
The first originates from supersymmetric gauge theories and the {\it point d'appui} is the syzygies of the vacuum moduli space as an algebraic variety; the second is key to arithmetical properties of algebraic varieties and whose foundations rest upon the Hasse-Weil zeta function.
  
\subsection{The Plethystic Programme for Gauge Theories}
Given a gauge theory, one of the most fundamental tasks is the construction of its gauge invariant operators.
In the case of supersymmetric gauge theories, especially those arising in the context of string theory, exhibiting as quiver gauge theories living on the world-volume of branes, due to the intrinsically geometrical nature of the situation, these operators are inextricably linked to the vacuum geometry of the theory.
We shall restrict our attention to $\cN=1$ supersymmetric gauge theories in $(3+1)$-dimensions on whose vacuum geometry we now briefly expound; the analysis extends itself readily to higher supersymmetries in other dimensions.

An $\cN=1$ (global) supersymmetric gauge theory is given by the action
\begin{equation}
S = \int d^4x\ \left[ \int d^4\theta\ \Phi_i^\dagger e^V \Phi_i +
    \left( \frac{1}{4g^2} \int d^2\theta\ \tr{W_\alpha W^\alpha} +
    \int d^2\theta\ W(\Phi) + {\rm h.c.} \right) \right] \ ,
\end{equation}
with the $\theta$ variables parameterizing $\cN=1$ superspace over which we integrate.
The $\Phi_i$ are chiral superfields transforming in some representation $R_i$ of the gauge group $G$; $V$ is a vector superfield transforming in the Lie algebra $\mathfrak{g} = Lie(G)$; $W_\alpha = i\overline{D}^2 e^{-V} D_\alpha e^V$, the gauge field strength, is a chiral spinor superfield; and $W(\Phi)$ is the superpotential, which is a holomorphic and typically polynomial, function of the $\Phi_i$.
Upon integrating over superspace, we obtain the scalar potential of the theory:
\begin{equation}
V(\phi_i, \bar{\phi_i}) = \sum_i \left| \diff{W}{\phi_i} \right|^2 + \frac{g^2}{4}(\sum_i q_i |\phi_i|^2)^2 \ ,
\end{equation}
where $\phi_i$ are the scalar components of the chiral fields $\Phi_i$, and with $q_i$ being their charges.
When the gauge group is Abelian, one could, in addition, allow a Fayet-Illiopoulos (FI) term to the Lagrangian: $\delta L = \int d^4 \theta \xi V \Rightarrow  (\sum\limits_i q_i |\phi_i|^2 - \xi)^2$ with FI parametre $\xi$.

\subsubsection{The Vacuum Moduli Space as an Algebraic Variety} \label{s:vms}
The vacuum of the above theory is the minimum of the potential, which, being a sum of squares, occurs when each squared quantity vanishes:
\begin{equation}\label{FD}
V(\phi_i, \bar{\phi_i}) = 0 \Rightarrow
\left\{
\begin{array}{cc}
\diff{W}{\phi_i} = 0 & \mbox{F-terms}\\
\sum\limits_i q_i |\phi_i|^2 = 0 & \mbox{D-terms}
\end{array}
\right. \ .
\end{equation} 
For supersymmetric theories, the (classical) vacuum, defined by the above flatness conditions for the F- and D-terms, is non-trivial, and is, in fact, an affine algebraic variety generically \cite{lt,Gray:2005sr,Gray:2006jb,VMS,witten93,Gray:2008yu} of greater than one complex dimension.
Therefore, there is a continuous moduli of vacua and this variety is commonly called the {\bf vacuum moduli space} (VMS), $\cM$.
The solution space of the F-terms alone is also of significant interest and is known as the {\bf master space} \cite{Forcella:2008bb,Forcella:2008eh} (cf. review in \cite{Forcella:2009bv}).

Whereas \eqref{FD} gives the explicit defining equation of $\cM$, the standard geometrical approach is to construe the D-terms as gauge invariants, subject to the vanishing of the F-terms. 
More specifically, $\cM$ is a GIT quotient of D-terms by the F-terms \cite{witten93,lt}.
Calculationally, one can establish a convenient algorithm \cite{Gray:2005sr,Gray:2006jb,Gray:2008yu}, using the techniques of computational algebraic geometry, to interpret the D-term invariants as a polynomial ring map from the ideal defined by the F-terms, with its image the ideal defining $\cM$ in a convenient Gr\"obner basis.
The procedure can be summarized succinctly as follows.
\begin{enumerate}
\item Let there be $m$ (scalar-components of super-)fields $\phi_{i=1,\ldots,m}$ and start with polynomial ring $\IC[\phi_1, \ldots, \phi_m]$;
\item Identify the obvious set of mesonic GIOs and find the generating set $D$, consisting of $k$ gauge invariants (the set of GIOs is always finitely generated); this prescribes a ring map between two polynomial rings:
 $\IC[\phi_1, \ldots, \phi_m] \stackrel{D}{\longrightarrow} \IC[D_1, \ldots, D_k]$;
\item Now incorporate superpotential $W$, generically a polynomial in the fields $\phi_i$ and find its Jacobian ideal of partial derivatives (F-flatness):
 $\langle f_{i=1,\ldots,m} = \diff{W(\phi_i)}{\phi_i} = 0 \rangle$;
\item The VMS is then explicitly the image of the ring map
    \[
    \frac{\IC[\phi_1, \ldots, \phi_m]} 
	 {\{ F = \langle f_1, \ldots, f_m \rangle \}}
    \stackrel{D=GIO}{\longrightarrow} 
    \IC[D_1, \ldots, D_k],
    \quad
    \cM \simeq {\rm Im}(D) \ . 
    \]
\end{enumerate}

The algebro-geometric nature of the VMS is especially pronounced in the context of string theory.
When a single brane is placed transversely to a non-trivial back-ground, as in the AdS/CFT correspondence, a duality is established between the world-volume physics and the bulk supergravity.
In particular, if a D3-brane is placed transverse to a local, affine Calabi-Yau three-fold singularity $\cM$, filling the 10-dimensions of type IIB, the world-volume is precisely a $(3+1)$-dimensional gauge theory with $\cN=1$ supersymmetry, product gauge group and bi-fundamental matter, encoded conveniently into a quiver.
By construction, the mesonic ({\it i.e.}, operators composed of direct contraction, involving no more than the Kronecker delta tensor, of the chiral fields) VMS of this gauge theory, computed from \eqref{FD}, is exactly the Calabi-Yau three-fold $\cM$.
When $N$ parallel coincident D3-branes are present, the VMS, due to the permutation on the branes, simply becomes $\mbox{Sym}^N \cM \simeq \cM^N / \Sigma_N$, the $N$-th symmetrized product of $\cM$.

\subsubsection{Gauge Invariants and the VMS}
We are interested in the complete spectrum of BPS mesonic operators in the gauge theory. As mentioned above, the space of these objects, quotiented by the F-flat constraints, should give rise to the VMS $\cM$.
These mesonic gauge invariant operators fall into two categories: single- and multi-trace. The former consists of words in the operators, with gauge-indices contracted but only a single overall trace and the latter, various products of the single-trace gauge invariants.
We shall denote the generating function of the single-trace gauge invariants at $N$ branes (or, equivalently, for $SU(N)$ matrices in the gauge theory) as $f_N(t; \cM)$ and that of the multi-trace invariants, $g_N(t; \cM)$; then, the $n$-th coefficient in the series expansion would enumerate the corresponding gauge invariants, where $n$ is a natural level corresponding to, for example, the total R-charge of the gauge theory.
Using the algebraic geometry of $\cM$ to construct $f$ and $g$, and hence to address the counting problem is the purpose of the so-called {\bf plethystic programme}, developed in \cite{Benvenuti:2006qr,Feng:2007ur} and furthered in
\cite{Forcella:2007wk,Butti:2007jv,Balasubramanian:2007hu,Hanany:2008qc,Gray:2008yu,Forcella:2008bb,Forcella:2008eh}.
Without much ado, let us briefly summarize the key points of this programme, referring the reader to details to {\it loc.~cit}.
\begin{itemize}
\item
The quantity $f_\infty(t)$, counting the single-trace gauge invariants at large $N$ is equal to $g_1(t)$; this is the {\it point d'appui} of our construction.
The level (R-charge) $n$  imposes a natural grading on the polynomial ring in which $\cM$ is an ideal and hence the generating function is the Hilbert series of $\cM$ in the suitable affine co\"ordinates embedded by the F- and D-terms:
\begin{equation}
f_{\infty}(t; \cM) = g_1(t; \cM) = HS(t; \cM) := 
\sum\limits_{n=0}^\infty a_n t^n \ .
\end{equation}
Here, the coefficient $a_n$ is simultaneously the number of single-trace gauge invariants at total R-charge $n$ and the complex dimension of the $n$-graded piece of the co\"ordinate ring prescribed by $\cM$;

\item
For arbitrary $N$, the relation between the single- and multi-trace GIOs are related by
\begin{equation}
g_1(t) = f_{\infty}(t); \qquad f_{\infty}(t) = PE[f_1(t)],
\quad g_{\infty}(t) = PE[g_1(t)]; \qquad g_{N}(t) = PE[f_N(t)]
\end{equation}
where $PE[~]$ is the {\bf plethystic exponential} functional defined as
\begin{equation}
f(t) = \sum\limits_{n=0}^\infty a_n t^n \quad \Rightarrow \quad
g(t) = PE[f(t)] = \exp\left( \sum_{n=1}^\infty \frac{f(t^n) -
  f(0)}{n} \right) =
\frac{1}{\prod\limits_{n=1}^\infty (1-t^n)^{a_n}} \ ;
\end{equation}
the structure of the infinite-product incarnation of this function should be reminiscent of the bosonic oscillator partition function;

\item
The plethystic exponentiation has an analytic inverse, called the plethystic logarithm and is given in terms of the number-theoretical M\"obius function $\mu(k)$:
\begin{eqnarray}
\nn
f(t) = PE^{-1}(g(t)) &=& \sum_{k=1}^\infty
\frac{\mu(k)}{k} \log (g(t^k)) \ , \\
\mu(k) &:=& \left\{\begin{array}{lcl}
0 & & k \mbox{ has repeated prime factors}\\
1 & & k = 1\\
(-1)^n & & k \mbox{ is a product of $n$ distinct primes} \end{array}\right.
\end{eqnarray}

\item
The defining equation, or syzygy, of $\cM$ is given by $f_1(t; \cM)$, which can be readily obtained from the plethystic logarithm of the Hilbert series:
\begin{equation}
f_1(t) = PE^{-1}[f_\infty(t)] = \mbox{ defining equation of $\cM$}.
\end{equation}
When, in particular, $\cM$ is a complete intersection variety, $f_1(t)$ is a terminating polynomial.

\item
To obtain the counting for arbitrary $N$, we promote $PE[~]$ to a parametre-inserted version and define the function
\begin{equation}
g(\nu ; t) := \prod\limits_{n=0}^{\infty} \frac{1}{(1
- \nu  \, t^n)^{a_n}} = \sum\limits_{N=0}^\infty g_N(t) \nu^N \ ,
\end{equation}
dependent on the fugacity parametre $\nu$.
The series expansion of $g(\nu ; t)$ in $\nu$ gives $g_N(t;\cM)$, the multi-trace generating function at given $N$, as its coefficients.
The single-trace generating function $f_N(t)$ is then retrieved as $PE^{-1}[g_N(t; \cM)]$.
\end{itemize}

We see, therefore, that the Hilbert series, through the plethystic functions, links the algebraic geometry of $\cM$ and the enumeration of the GIOs, single- and multi-traced, of the $\cN=1$ gauge theory whose VMS is $\cM$.
It is thus expedient to quickly remind the reader some key features of $HS(t;\cM)$.
It is important that the Hilbert series is a {\it rational function} in $t$ and can be expressed in two ways:
\begin{equation}
H(t; {\cal M}) = \left\{
  \begin{array}{ll}
  \frac{Q(t)}{(1-t)^k} \ , & \mbox{ Hilbert series of the first kind} ~;\\
  \frac{P(t)}{(1-t)^{\dim({\cal M})}} \ , & \mbox{ Hilbert series of the
    second kind} \ ,  \end{array} \right.
\end{equation}
where $k$ is the dimension of the ambient affine space in which $\cM$ embeds.
In either guise, the numerators $P(t)$ and $Q(t)$ are polynomials with {\it integer} coefficients and $P(1)$ is the degree of $\cM$.
The powers of the denominators are such that the leading pole captures the dimension of the embedding space and the manifold, respectively.
When expanding the Hilbert series into a Taylor series, then, the coefficients have polynomial growth and is called the Hilbert polynomial. 
We remark that for {\it non-commutative} graded rings and associated spaces, it is not necessary that the Hilbert series be rational \cite{NChilb}; it would indeed be interesting to investigate such circumstances, especially since to D-brane gauge theories one should associate a non-commutative algebra, whose centre is the classical VMS \cite{Berenstein:2002ge}.

\subsection{Algebraic Varieties over Finite Fields}\label{s:zeta}
In this section, we remind the reader of some well-known results on algebraic varieties defined over finite fields.
In particular, we shall address some basic principles of local zeta functions and the Weil Conjectures as well as global zeta function and their manifestation as L-functions. 
In due course, we will see expressions reminiscent of our plethystic functions discussed in the previous section.

First, by a finite field we mean a number field with a finite number of elements (cf.~\cite{serre}).
We are acquainted with $\IF_p$; this is simply the cyclic group of $p$ elements for some prime number $p$ (called the {\bf characteristic} of the field), with members represented by $0,1,2,\ldots,p-1$ and the field axioms for addition, multiplication and division being the usual arithmetic modolo $p$.
Less familiar are perhaps the finite (Galois) extensions of this field, these are $\IF_{p^r}$, consisting of $p^r$ elements for some positive integer $r$.
These elements can be explicitly constructed as follows.
Take the polynomial ring $\IF_p[T]$ with coefficients in the field $\IF_p$, and consider a monic irreducible polynomial $f(T)$ of degree $r$; then the quotient ring, by the principal ideal $(f(T))$ $\IF_p[T] / (f(T))$, is a field of $p^r$ elements.
It is well-known that $\IF_{p^r}$ for various primes $p$ and positive integers $r$ are all the finite number fields.

Our main purpose is to consider an algebraic variety $X$, defined not over the field $\IC$ of complex numbers, but, rather, over $\IF_{p^r}$, {\it i.e.}, all indeterminates are to take values in these finite fields ({\it cf.~e.g.} \cite{hartshorne,modCY,Candelas:2004sk}).
As such, our usual notion of a variety now simply becomes a discrete set of points.
An exponentiated generating function, can be formed for the number of points $N_{p^r}$ of $X$ over $p^r$:
\begin{equation}\label{zeta}
Z_p(t) = \exp\left( \sum\limits_{r=1}^\infty \frac{N_{p^r}}{r} t^r \right) \ ;
\end{equation}
this is dubbed the {\bf local zeta function} of the variety $X$ for the prime $p$ because it is localized at a fixed prime.

We can delocalize by taking the product of the above over all primes (we will encounter so-called primes of good versus bad reduction later)
giving us the so-called {\bf global}, or {\bf Hasse-Weil}, zeta function:
\begin{equation}
Z(t) = \prod\limits_p Z_p(t) \ .
\end{equation}
It is customary to apply the substitution $t := p^{-s}$ to the above, for reasons which will soon become apparent in the ensuing examples, to which we now turn.

\paragraph{A Point: }
The simplest variety is a single point.
In this case, $N_{p^r} = 1$ for all $p$ and all $r$. Whence, 
\begin{eqnarray}
Z_p(t; \mbox{pt}) &=& \exp\left( \sum\limits_{n=1}^\infty \frac{t^r}{r} \right) = \frac{1}{1-t} \ ,\\
\label{euler}
Z(s; \mbox{pt}) &=& \prod\limits_p \frac{1}{1 - p^{-s}} = \zeta(s) \ ,
\end{eqnarray}
where, in the second line, we have used the standard Euler product for the Riemann zeta function.
This illustrative example indeed should clarify the various names and substitutions stated above.

\paragraph{The Projective Space $\IP^k$: }
For the case of the affine line $\IA^1$ over $\IF_{p^r}$, there are clearly $p^r$ points; the projectivization introduces a point at infinity, and hence $N_{p^r} = p^r + 1$, giving us
\begin{eqnarray}
Z_p(t; \IP^1) &=& \exp\left( \sum\limits_{r=1}^\infty \frac{p^r+1}{r}t^r \right) = \frac{1}{(1-t)(1-p t)} \ ,\\
\label{zeta-P1}
Z(s; \IP^1) &=& \prod\limits_p \frac{1}{(1 - p^{-s})(1 - p^{-s+1})} 
  = \zeta(s)\zeta(s-1) \ .
\end{eqnarray}
The affine line itself would have simply given $Z_p(t; \IA^1) = \frac{1}{1-pt}$ and $Z(s; \IA^1) = \zeta(s-1)$.
The generalization to $\IP^k$ over ${\IF_{p^r}}$ is straight-forward.
Recall that $\IP^k \sim (\IA^{k+1} \setminus \{0,0,\ldots,0\}) / \IF_{p^r}^*$ where $\IF_{p^r}^*$ are the non-zero elements in the field, totalling $p^r-1$ in number. 
Thus, the number of points in $\IP^k$ is $(p^r)^{k+1} - 1$ quotiented by $p^r-1$, giving us $N_{p^r} = (p^{r(k+1} - 1) / (p^r-1) = 1 + p^r + p^{2r} + \ldots + p^{kr}$.
Therefore,
\begin{eqnarray}
\label{zetaPN}
Z_p(t; \IP^k) &=& \exp\left( \sum\limits_{r=1}^\infty 
\frac{\sum\limits_{j=0}^k p^{k j}}{r}t^r \right) 
  = \prod_{j=0}^k \frac{1}{(1- p^j t)} \ ,
\\
Z(s; \IP^k) &=& \prod\limits_p \prod_{j=0}^k \frac{1}{(1- p^{j-s})}
  = \prod_{j=0}^k \zeta(s-j) \ .
\end{eqnarray}
Similarly, we have that, for $\IC^k$, $N_n = p^{kn}$, giving us
\begin{equation}\label{zetaCN}
Z_p(t; \IC^k) = (1-p^k t)^{-1}, \qquad Z(s; \IC^k) = \zeta(s-k) \ .
\end{equation}

\subsubsection{The Dirichlet Series}\label{Lseries}
The Euler product in \eqref{euler}, a key property for the Riemann zeta function, is a general feature of the global zeta functions of our concern.
Indeed, one can develop an expansion of such products, into what is known as an {\bf Dirichlet Series} (sometimes called L-series for reasons which shall become clear later):
\begin{equation}
Z(s) = \prod\limits_p Z_p(p^{-s}) = \sum\limits_{n=1}^\infty \frac{c_n}{n^s} \ ,\end{equation}
where the product over primes is naturally converted to a sum over the integers. For the above example of a single point, the coefficients $c_n$ are simply all unity, the famous representation of $\zeta(s)$.

Some standard identities in the theory of Dirichlet series (cf.~\cite{dirichlet}) will be pertinent to us.
The first is that, for $\re(s) > \max(1, 1 + \re(a))$,
\begin{equation}\label{zz-dir}
\zeta(s)\zeta(s-a) = \sum\limits_{n=1} \frac{\sigma_a(n)}{n^s} \ ,
\end{equation}
where we recall that $\sigma_a(n) := \sum\limits_{d | n} d^a$ is the sum over the $a$-th power of the divisors of $n$.
This is a special case of the so-called convolution property of Dirichlet series, that
\begin{equation}
A(s) = \sum_{n=1} \frac{a_n}{n^s} \ ,
B(s) = \sum_{n=1} \frac{b_n}{n^s} \quad \Rightarrow
A(s)B(s) = \sum_{n=1} \frac{(a*b)_n}{n^s} \ , 
\end{equation}
with the convolution defined as
\begin{equation}
(a*b)_n = \sum\limits_{i | n} a_i \ b_{\frac{n}{i}} \ .
\end{equation}

In general, the coefficients of the L-series can be obtained by transform of Mellin-type. This inversion is the so-called Perron's formula and states that for $Z(s) = \sum\limits_{n=1}^\infty \frac{c(n)}{n^s}$, convergent when $\re(s) > a$,
\begin{equation}
A(m) = \sum_{n=1}^{m-1} c(n) + \frac12 c(m) 
= \frac{1}{2\pi i} \int_{c-i\infty}^{c+i\infty} Z(z) m^z \frac{dz}{z} \ ,
\end{equation}
for positive integer $m$, $c$ an arbitrary real number such that $\re(s) > a - c$.
These partial sums can then be listed to iteratively obtain the individual terms desired. 
We also have a direct, though less computationally straight-forward inversion formula that
\begin{equation}
c(n) = \lim\limits_{T \rightarrow \infty} \frac{1}{2T} \int_{-T}^{T}
Z(s) n^s d(\im(s)) \ .
\end{equation}
Finally, one could use an even more direct method of taking limits, wherever possible.
Indeed, writing, by setting $s:= -\log x$ so that as $x$ tends to 0, $s$ tends to infinity,
\begin{equation}
Z(s) = 
\sum\limits_{n=1}^\infty \frac{c_n}{n^s} = 
\sum\limits_{n=1}^\infty c_n n^{\log x} \ ,
\end{equation}
we have that $c_1 = \lim\limits_{x \to 0} Z(-\log x)$, whence 
$c_2 = \lim\limits_{x \to 0} 2^{-\log x}(Z(-\log x) - c_1)$, and so on, successively, until order by order all the coefficients are obtained.

\subsubsection{Weil-Grothendieck-Deligne}\label{s:weil}
The general structure of the local zeta function of the algebraic variety $X$ reflects, in a remarkably elegant fashion, the geometrical nature of $X$.
These are captured by what historically have come to be known as the {\bf Weil Conjectures} (1940's) and proved by Deligne in 1974, using the $\ell$-adic cohomological techniques envisioned by Grothendieck.
In summary, the statement is that for $X$ a (non-singular) $m$-dimensional complex projective algebraic variety,
\begin{itemize}
\item 
The local zeta function $Z(t; X)$ is a rational function which can be more precisely written as alternating products in numerator and denominator:
\begin{equation}\label{Zp}
Z_p(t; X) = \frac{P_1(t) \cdots P_{2m-1}(t)}{P_0(t) \cdots P_{2m}(t)} \ ,
\end{equation}
where $P_0(t) = 1-t$, $P_{2m}(t) = 1 - p^m t$ and $P_{i=1,\ldots, 2m-1}$ are all polynomials with integer coefficients, admitting factorization $\prod\limits_j (1 - \alpha_{ij} t)$ with $\alpha_{ij}$ being complex numbers with modulus $|\alpha_{ij}| = p^{\frac{i}{2}}$. When $m=1$, this means that all zeros of $Z_p(t; X)$ are at $\re(s) = \frac12$, an analogue of the Riemann Hypothesis.

\item
A functional equation $Z_p(\frac{1}{p^m t}; X) = \pm p^{\frac{m \chi}{2}} t^{\chi} Z_p(t; X)$, where $\chi$ is the Euler number of $X$, is obeyed.

\item
For primes $p$ of good reduction (that is, when the variety remains non-singular over ${\IF_{p^r}}$, a point which we discuss in Appendix \ref{A:reduction}), the degree of $P_i$ is the $i$-th Betti number of $X$ as a complex variety.
\end{itemize}

\section{Exempli Gratia}\label{s:eg}
Having given a brief account of our two protagonists, it is now expedient to present, in relation thereto, some illustrative examples of gauge theories, their moduli spaces and associated arithmetic as well as geometric properties, as much as a warm-up, as a provision of a small catalogue against which one could initiate some systematic checks and experiments (note that our approach is different from {\it e.g.}\cite{Nambu:1985tw,connes} to whose penetrating insights the reader is highly encouraged to refer).

\subsection{A Single Field}\label{s:1field}
The simplest gauge theory is undoubtedly that of $SU(N)$ with a single free field $X$ charged therein, embodied as $N \times N$ matrices. There is no superpotential and the gauge invariants are simply $\tr X^i$ for $i = 1, \ldots, N$, with  any power $i>N$ re-writable, in terms of Newton polynomials, in this fundamental generating set. 
When $N\to \infty$ where no such matrix relations occur, we have that the moduli space is simply the affine complex line $\IC$, generated by $\tr(X)$ and the full specturm of operators are $\II, \tr(X), \tr(X^2), \tr(X^3), \ldots$ 
Thus, the fundamental generating function is the Hilbert series for $\IC$ (cf.~\S 7.1 of \cite{Benvenuti:2006qr}):
\begin{equation}
g_1(t; \IC) = f_\infty(t; \IC) = \frac{1}{1-t} = 1 + t + t^2 + t^3 + \ldots
\end{equation}

The plethystic logarithm gives $PE^{-1}[f_\infty(t;\IC)] = f_1(t;\IC) = t$, signifying precisely the above: that the chiral ring is freely generated by a single element. 
The plethystic exponential is the Euler Eta-function $\prod\limits_{n=1}^\infty (1-t^n)^{-1}$, whose expansion encodes the (free) partition of integers and corresponds to the various ways the above single-trace operators can be multiplied. 
Finally, the $\nu$-inserted plethystic exponential gives
\begin{equation}
g_\nu(t; \IC) = \prod\limits_{m=0}^\infty (1 - \nu t^m)^{-1} = 
\sum\limits_{N=1}^\infty \prod\limits_{n=1}^N \frac{1}{1-t^n} \nu^N \ ,
\quad
f_N(t;\IC) = PE^{-1}[g_N(t;\IC)] = \frac{1-t^{N+1}}{1-t} \ .
\end{equation}
Therefore, at fixed $N$, the VMS becomes the symmetric product $\IC^N / \Sigma_N$, and the corresponding single- and multi-trace spectra are counted, respectively, by $f_N(t;\IC)$ and $g_N(t; \IC)$.
Strictly speaking, one should think of the VMS as being the Hilbert scheme of $N$-points on $\IC$ (q.~v.~\cite{nakajima}) and plethystics for these are discussed in \cite{Feng:2007ur}.

The arithmetic properties of the above VMSs are also readily computed.
For $\IC$, the local and global zeta functions were given in \eqref{zetaCN}:
\begin{equation}\label{zetaC1}
Z_p(t; \IC) = \frac{1}{1 - p \ t} \ , \quad
Z(s; \IC) = \zeta(s-1) \ .
\end{equation}
For $N > 1$, the spaces become more involved, even though their Hilbert series are neatly compacted into $f_N(t; \IC)$.
Let us begin with $N=2$.
Here we have the affine variety described by $(x,y) \leftrightarrow (y,x)$ acting on $\IC[x,y]$. The Hilbert series is given by $g_2(t; \IC) = [(1-t)(1-t^2)]^{-1}$, signifying a freely generated algebra by two elements, of degrees 1 and 2, respectively: these we see clearly as the invariants $x+y$ and $x^2+y^2$ under the exchange.
The VMS is therefore an affine cone over the weighted projective space $W\IP^1_{[1:2]}$.
Thus, other than the weighting of the c\"oordinates, the variety is simply $\IC^2$ and no extra syzygies exist for our particular embedding.
Hence, the zeta functions are simply $Z_p(t; \IC^2) = (1 - p^2 \ t)^{-1}$ and $Z(s; \IC^2) = \zeta(s-2)$.
This treatment generalizes to arbitrary $N$, giving the VMS as the affine cone over $W\IP^1_{[1:2:\ldots:N]}$, isomorphic to $\IC^N$, and having zeta functions as given in \eqref{zetaCN}.

Parenthetically, the projectivization of the above is also interesting. Adding a point at infinity gives $\IP^1$, and the global zeta function, from \eqref{zeta-P1}, is the product $\zeta(s) \zeta(s-1)$. Consequently, the L-series is
\begin{equation}
Z(s; \IP^1) = \zeta(s)\zeta(s-1) = \sum\limits_{n=1}^\infty \frac{\sigma_1(n)}{n^s} \ ,
\end{equation}
where we have used \eqref{zz-dir}.
If we were to develop the power series with the L-series coefficients $c_n = \sigma_1(n)$, we would have $g(t) = 1+\sum\limits_{n=1} \sigma_1(n) t^n$.
This can be re-written  \cite{dirichlet} as a so-called {\bf Lambert summation}, $\sum\limits_{n=1}^\infty \sigma_a(n) t^n = \sum\limits_{n=1}^\infty \frac{n^a t^n}{1 - t^n}$ where the number theoretic divisor function becomes implicit.

\subsection{D3-Brane in Flat Space}
One of the most studied gauge theory in recent times is unquestionably the world-volume theory of $N$ parallel coincident D3-branes, especially in the context of holography and Maldacena's AdS/CFT Correspondence.
The simplest setup is that of the D3-brane transverse to flat $\IC^3$, considered as a real cone over $S^5$, with near-horizon geometry of $AdS_5 \times S^5$.
The world-volume theory is $\cN=4$, $U(N)$ super-Yang-Mills theory in 4-dimensions with three adjoint fields, say $x,y,z$, charged under the $U(N)$.
There is a simple cubic superpotential $W = \tr(x[y,z])$ and the matter content can be easily represented by the clover quiver:
\begin{equation}
\begin{array}{c}\includegraphics[trim=0mm 0mm 0mm 0mm, clip, width=1in]{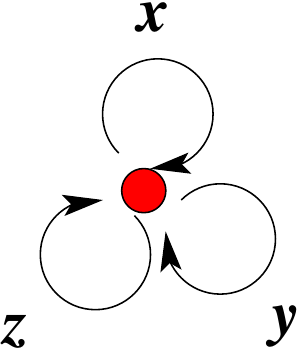}
\end{array}
\qquad
\qquad
W = \tr(x[y,z]) \ ,
\end{equation}
where the node corresponds to the $U(N)$ gauge group and the three arrows, the three (adjoint) fields.

The VMS, by construction, is parameterized by the transverse motion of the branes and subsequently is $\IC^3$ for a single D3-brane and the $N$-th symmetrized product thereof for arbitrary $N$.
The F-terms, from the Jacobian of $W$, demand that $x,y,z$ mutually commute and we have the symmetric commutative algebra generated by three elements.
The plethystics were computed in \cite{Benvenuti:2006qr} and, with the standard binomial symbol ${x \choose 2}$ we have
\begin{eqnarray}\label{plethconi}
&&g_1(t; \IC^3) = f_\infty(t; \IC^3) = \frac{1}{(1-t)^3} = 
\sum\limits_{n=0}^\infty {n + 2 \choose 2} t^n \ , \\
\nn
&&
g_\nu(t; \IC^3) = \prod\limits_{m=0}^\infty (1 - \nu t^m)^{-\frac{(n+2)(n+1)}{2}} = 
\sum\limits_{N=1}^\infty g_N(t; \IC^3) \nu^N \ ,
\quad
f_N(t;\IC^3) = PE^{-1}[g_N(t;\IC^3)] \ .
\end{eqnarray}

Now, for $N=1$, we have $\IC^3$ and the zeta functions are back to \eqref{zetaCN}. At $N=2$, expanding the $\nu$-inserted plethystic exponential gives us
\begin{equation}
g_2(t; \IC^3) = \frac{1+3t^2}{(1-t)^3(1-t^2)^3} \ , \quad
f_2(t; \IC^3) = 3t + 6t^2 - 6t^4 + 8t^6 - 18t^8 + \cO(t^{10}) \ .
\end{equation}
The non-terminating syzygies in $f_2$ signifies that we have a VMS which is non-complete intersection, whose Hilbert series is given by $g_2$.
Luckily, because of the relative simplicity of the space, we can readily write down the invariants. Let $x,y,z$ be the c\"oordinates of $\IC^3$ (this is not an abuse of notation, the three fields, after imposing the commuting F-terms, should correspond precisely to these affine c\"oordinates), then our $\Sigma_2$ action takes these to, say $x',y',z'$, and we have the full 6 c\"oordinates of $(\IC^3)^2$.
In degree one, the invariants are clearly $u_1 = x+x'$, $u_2 = y+y'$ and $u_3 = z+z'$.
In degree two, the invariants are the obvious $v_1 = xx'$, $v_2 = yy'$ and $v_3 = zz'$, as well as $v_4 = x y + x'y'$, $v_5 = x z + x'z'$ and $v_6 = y z + y'z'$.
Since $\Sigma_2$ is a group of order 2, by N\"other's theorem on invariants, we need not look for higher invariants.
We can then calculate, facilitated by the aid of \cite{m2}, that there are non-trivial syzygies amongst these invariants: we obtain the non-complete intersection which is an affine complex cone over 6 quartics in $W\IP^9_{[1:1:1:2:2:2:2:2:2]}$. Explicitly, the equations are
\begin{eqnarray}
\nn
(\IC^3)^2 / \Sigma_2 
&\simeq& 
\{u_2^2 v_3-u_3 u_2 v_6+u_3^2 v_2+v_6^2-4 v_2 v_3 \ ,
u_1^2 v_3-u_3 u_1 v_5+u_3^2 v_1+v_5^2-4 v_1 v_3 \ , \\
\nn &&
u_1^2 v_2-u_2 u_1 v_4+u_2^2 v_1+v_4^2-4 v_1 v_2,\\
\nn &&
-u_3^2 v_4-u_2 u_3 v_5-u_1 u_3 v_6-2u_1u_2v_3+
   u_1 u_2 u_3^2+4 v_3 v_4+2 v_5 v_6,\\
\nn &&
-u_2^2 v_5-u_3u_2v_4-u_1u_2v_6-2 u_1 u_3 v_2+u_1 u_3 u_2^2+
   4 v_2 v_5+2 v_4 v_6,\\
 &&
-u_1^2 v_6-u_3 u_1v_4-u_2 u_1 v_5-2 u_2 u_3 
v_1+u_2 u_3 u_1^2+2v_4v_5+4v_1 v_6
\} \ .
\end{eqnarray}
Given this algebraic variety, we can proceed to compute its zeta function.
The base of the cone is a complex 5-dimensional projective variety , therefore
the local zeta function, by \eqref{Zp}, should be of the form
$\frac{P_1(t)P_3(t)P_5(t)P_7(t)P_9(t)}
 {(1-t) P_2(t) P_4(t) P_6(t) P_8(t)(1-p^5t)}$, where $P_i(t)$, for $i=1,\ldots,9$, is a polynomial of degree $b_i$, the $i$-th Betti number of the 10 real-dimensional base manifold.
Now, the cone has one more point than the base, {\it viz.}~the tip at, say, the origin, which is removed when projectivizing.
Thus, this addition of unity to $N_{p^r}$, upon exponentiating according to the definition \eqref{zeta}, gives a trivial factor of $\exp(\sum\limits_{r=1}^\infty \frac{t^r}{r}) = (1-t)^{-1}$. 
Hence, the total affine variety has zeta-function
\begin{equation}
Z_p(t; (\IC^3)^2 / \Sigma_2 ) = 
\frac{P_1(t)P_3(t)P_5(t)P_7(t)P_9(t)}
 {(1-t)^2 P_2(t) P_4(t) P_6(t) P_8(t)(1-p^5t)} \ .
\end{equation}
The coefficients of these polynomials can be fixed by tabulating the explicit number of solutions for some low values of $p^r$.
We find, for example, that $N_{p^1} = 2^6, 3^6, 5^6$ for the first few primes, a pattern which we speculate will persist to hold.
For now, let us not belabour the point and turn to demonstrate the determination of such coefficients for some simpler examples.

\subsection{The Conifold}
A relatively simple gauge theory, canonical in the string theory literature \cite{Klebanov:1998hh}, is the theory of D3-branes on a conifold $\cC$, {\it i.e.}, the quadric in $\IC^4$ or $\{u v = z w \} \subset  \IC[u,v,z,w]$.
Note that this is a toric variety and enumeration of gauge invariants are greatly facilitated thereby; we include the toric diagram below, drawn in a plane due to the Calabi-Yau nature, for reference.
The world-volume theory has $\cN=1$ supersymmetry, $SU(N) \times SU(N)$ gauge group, with four bi-fundamental fields as well as a quartic superpotential:
\begin{equation}\begin{array}{ll}
\begin{array}{c}
\includegraphics[trim=0mm 0mm 0mm 200mm, clip, width=3in]{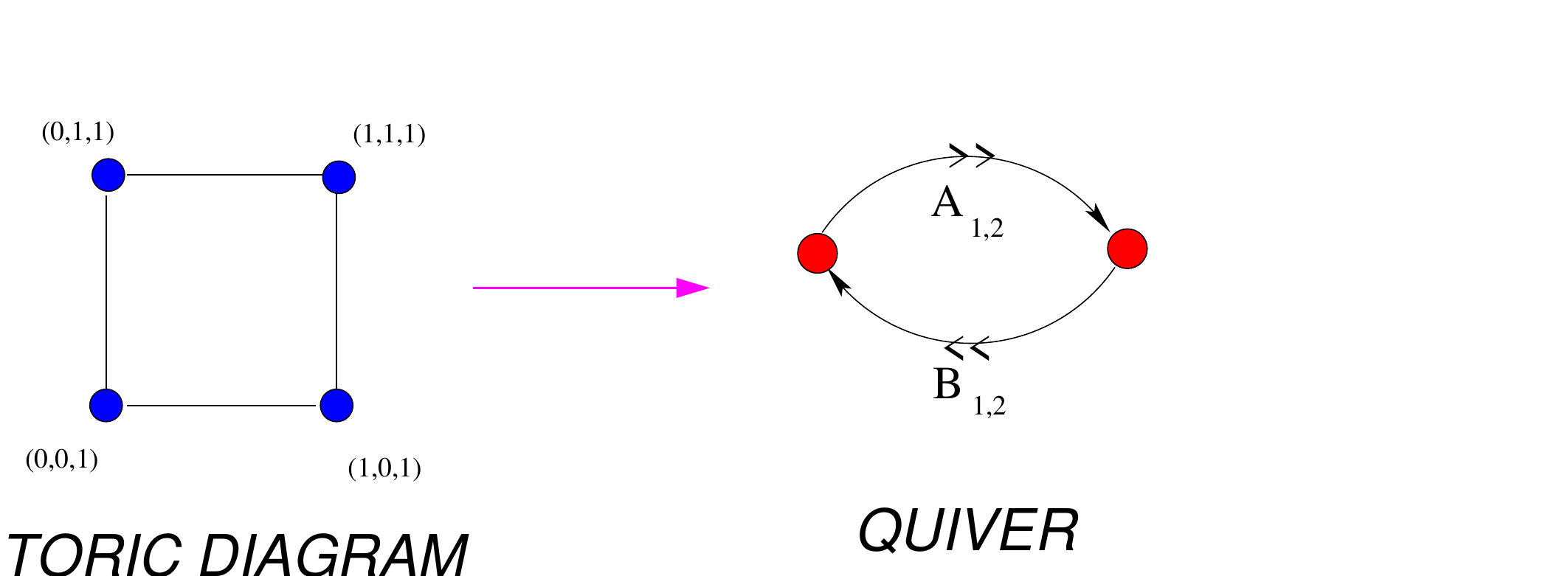}
\end{array}
&
{\small
\begin{array}{l}
\begin{array}{ccc}
    & SU(N) & SU(N) \\
A_{i=1,2} & \fund & \antifund \\
B_{j=1,2} & \antifund & \fund
\end{array}\\
W = \tr(\epsilon_{il} \epsilon_{jk} A_i B_j A_l B_k)
\end{array}}
\end{array}\end{equation}

The counting of the gauge invariants can be done explicitly \cite{Benvenuti:2006qr,Feng:2007ur}. We have four fundamental invariants, corresponding to the four Euler cycles in the quiver \cite{Hewlett:2009bx}, $M_{0,1}= A_1 B_1,\quad M_{1,0}=A_1 B_2,\quad M_{-1,0}=A_2 B_1,\quad M_{0,-1}=A_2 B_2$, subjecting to the F-term relation obtained from the quartic superpotential: $M_{0,1} M_{0,-1}= M_{1,0}M_{-1,0}$.
Diagrammatically, we can then see the lattice points in the toric cone corresponding to the gauge invariants (cf.~\cite{balazs}):
\[
\begin{array}{c}
\includegraphics[trim=0mm 0mm 0mm 0mm, clip, width=6in]{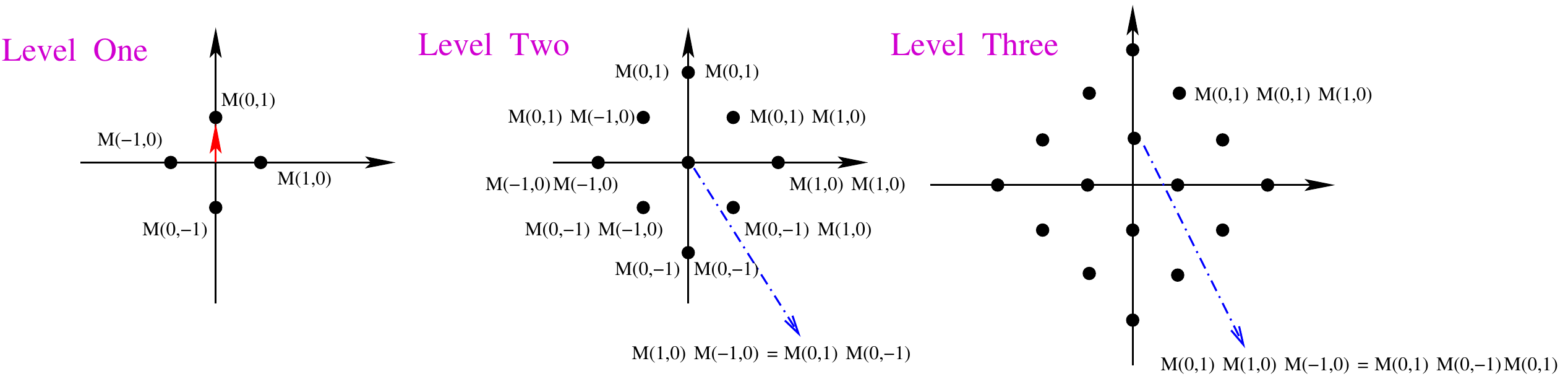}
\end{array}
\]
Subsequently, the VMS is by construction the cone over the said quadric hypersurface for a single D-brane and for arbitrary $N$, plethystic analysis gives us:
\begin{eqnarray}
&&
\nn
g_1(t; \cC) = f_\infty(t; \cC) = \frac{1+t}{(1-t)^3} = 
\sum\limits_{n=0}^\infty (n+1)^2 t^n \ , \\
\label{fCon}
&&g_\nu(t; \cC) = \prod\limits_{m=0}^\infty (1 - \nu t^m)^{(n+1)^2}
= 
\sum\limits_{N=1}^\infty g_N(t; \IC^3) \nu^N \ .
\end{eqnarray}
We see that the $(n+1)^2$ indeed captures the lattice cone counting above.

Now, we have a rather simple projective variety, a single hyper-surface, call it $C$, of degree 2 in $\IP^3$, over which $\cC$ is an affine complex cone.
\comment{
The counting of points $N_{p^r}$ over finite fields for $\cC$ is that of $C$, plus the one point at the origin where the apex of the cone resides.
Thus, their local zeta function differ by a factor of the form $\exp(\sum\limits_{n=1}^\infty\frac{1}{n}t^n) = (1-t)^{-1}$:
\begin{equation}\label{Zpconi}
Z_p(t; \cC) = \frac{1}{1-t} Z_p(t; C) \ .
\end{equation}
}
This base surface $C$ is clearly a K\"ahler manifold and the Hodge diamond is
$h^{0,0} = 1$, $h^{0,1}=h^{1,0}=0$, $h^{0,2}=h^{2,0}=0=h^{1,2}=h^{2,1}$, $h^{1,1}=2$ and $h^{2,2}=1$.
Whence, the Betti numbers\footnote{The reader versed in AdS/CFT is perhaps more used to the number of 2-cycles being 1, thinking of the conifold as a real cone over $S^2 \times S^3$. Here, however, we are considering it as a complex cone over the quadric surface and will study this compact base surface here.}
are $b_0 = b_4 = 1$, $b_1 = b_3 = 0$, $b_2 = 2$.
Again, because $\cC$ has one more point, at the tip of the cone, than $C$, the local zeta function is as dictated by \eqref{Zp}, together with an additional factor of $(1-t)^{-1}$:
\begin{equation}\label{ZpC}
Z_p(t; \cC) = \frac{1}{(1-t)^2(1 - A_p t + pt^2)(1-p^2t)} \ .
\end{equation}
In the denominator, the only non-trivial factor would have been $P_2(t)$, a quadratic form which we have spelt out, wherein a single indeterminate, {\it viz.}, $A_p$ is to be fixed.
It therefore suffices to enumerate at $p^{r=1}$ to determine $A_p$ and govern all the finite Galois extensions thereof in a single sweep.

Comparing \eqref{ZpC} with the definition for $Z_p$ and factorizing $1 - A_p t + pt^2 = (1-\alpha t)(1-\beta t)$,  we find that
\begin{eqnarray}
\nn
\sum\limits_{r=1}^\infty \frac{N_{p^r}}{r} t^r 
&=& \log Z_p(t; \cC) =
    -2\log(1-t) - \log(1-p^2t) - \log(1-\alpha t) - \log(1-\beta t)\\
&=& 
\sum\limits_{r=1}^\infty \frac{t^r}{r} \left[ 2 + p^{2r} + \alpha^r + \beta^r
\right] \ . 
\end{eqnarray}
Whence, we can determine the coefficient $A_p$ as
\begin{equation}\label{NpAp}
N_{p^r} = 2 + (\alpha^r + \beta^r) + p^{2r} \Rightarrow
A_p = \alpha+\beta = N_{p^{r=1}} - 2 - p^2 \ .
\end{equation}
We can readily find the first values of $N_p$ on the computer:
\begin{equation}\label{Npconi}
N_{p^{r=1}} = \{ 10, 33, 145, 385, 1441, 2353, 5185, 7201, 12673, 25201 \ldots \} \ ;
\end{equation}
interestingly, these are all square-free integers:
\[
\{{2\cdot5}\ , {3\cdot11}\ , {5\cdot29}\ , {5\cdot7\cdot11}\ ,
{11\cdot131}\ , {13\cdot181}\ , {5\cdot17\cdot61}\ , \
{19\cdot379}\ , {19\cdot23\cdot29}\ , {11\cdot29\cdot79}\} \ .
\]
From these we can determine the global zeta function, and thence its L-series development:
\begin{eqnarray}
\nn
Z(s; \cC) &=&  \prod_{p} \frac{1}{(1-p^{-s})^2}
 \frac{1}{(1-p^{2-s})(1+p^{1-2s}+p^{-s}(N_p-1-p^2))} =
 \sum\limits_{i=1}^\infty \frac{c_n}{n^s} \ , \\
&& c_n = \{ 1, 11, 34, 80, 146, \ldots \} \ .
\end{eqnarray}

What is perhaps more interesting is, upon seeing no immediate pattern to the above, when one desingularizes the cone by a standard deformation of complex structure. In particular, let us consider the variety $u v - z w = 1$, which we shall denote $\widetilde{\cC}$. 
We have chosen the complex parametre to be 1 to avoid it being reduced back to 0 for some prime factor, constituting an obviously bad reduction.
\comment{
Now, because we have removed the apex of the cone at the origin, the form of \eqref{ZpC} is the full result of the conifold, without the $(1-t)^{-1}$ factor.}
In this case, we find that
\[
N_{p^{r=1}} = \{ 6, 24, 120, 336, 1320, 2184, 4896, 6840, 12144, 24360, \ldots \} \ ,
\]
or, as one could emperically convince oneself, $N_{p} = p(p^2-1)$.
This gives us the form of the local zeta function, using \eqref{ZpC}, as
\begin{equation}
Z_{p}(t; \widetilde{\cC}) = \frac{1}{(1-t)^2} \frac{1}{(1-p^2 t)}
\frac{1}{(1+pt^2-(p^3 - p^2 - p - 1)t)} \ ,
\end{equation}
and subsequently the global zeta function as
\begin{equation}\label{zetaConifold}
Z(s; \widetilde{\cC}) = \zeta(s)^2 \zeta(s-2)
\prod_p (1+p^{1-2s}+(p^3 - p^2 - p - 1)p^{-s})^{-1}
\ .
\end{equation}
The function in the form of the infinite product can be thought of as an L-function for this variety; we shall return to this type of function for a more canonical example, involving the elliptic curve, later on in \S\ref{s:E} and Appendix \ref{A:reduction}.
Using the form of \eqref{zetaConifold}, we can also perform a Dirichlet expansion to obtain:
\begin{eqnarray}
\label{DirichletC}
&&Z(s; \widetilde{\cC}) =  \sum\limits_{n=1}^\infty \frac{c_n}{n^s} \\
&&
\nn
c_n = 
\{
1, 7, 25, 32, 121, 175, 337, 130, 449, 847, 1321, 800, 2185, 2359,
3025, 519, 4897, 3143, \ldots \}
\end{eqnarray}

Now, for higher $N$, the space becomes more involved.
Take $N=2$, for a brief example.
We have a $\Sigma_2$ action on $\IC^8[u,v,z,w,u',v',z',w']$, exchanging the primed and unprimed c\"oordinates, and in addition we have two copies of the defining quadric equations.
The invariants are $u+u'$, $v+v'$, $z+z'$ and $w+w'$ at degree 1, $uu'$, $vv'$, $zz'$, $ww'$, as well as the combinations $uv+u'v'$, $uz+u'z'$, $uw+u'w'$, $vz+v'z'$, $vw+v'w'$ and $zw+z'w'$ at degree 2, which should be subject to $uv-zw=0$ and $u'v'-z'w'=0$.
This is subsequently gives a non-complete-intersection, of complex dimension 6, defined by four cubics and twenty quartics, embedded in $W\IP^{14}_{[1^4:2^{10}]}$.
The Hilbert series can be computed, either from \cite{m2}, or from \eqref{plethconi}, to be $f(t; \cC^2 / \Sigma_2) = g_2(t; \cC) = \frac{4 t^4+3 t^3+7 t^2+t+1}{(t-1)^6 (t+1)^3}$.

Of parenthetical interest is perhaps the master space in the toric, or $N=1$, case. This space is the solution set to the F-terms alone and controls, in the sense of GIT quotient, the final VMS; it has been extensively studied for gauge theories in \cite{Forcella:2008bb,Forcella:2008eh}.
For the present case of the conifold, the space is simply $\IC^4$ and the zeta functions once more return to the simple form in \eqref{zetaCN}.

As a further digressive remark, we know that the gauge theories for the so-called generalized conifold $uv = z^m w^k$ and orbifolded conifold $\{uv = y^m, zw=y^k\}$ have been extensively studied (cf.~\cite{genconi}). Reducing these varieties  over some low primes, however, produced the same enumeration as of points as did the above for the conifold itself. This should be due to the Frobenius automorphism from points to its power, reduced over characteristic $p$ and hence produce no further points.

\subsection{Abelian Orbifolds of $\IC^3$}
It was shown in \cite{Feng:2000mi} that since any toric Calabi-Yau threefold singularity can be obtained from partial resolutions of $\IC^3/\IZ_k^2$ for sufficiently large $k$, all toric gauge theories on D3-branes can be algorithmically obtained by Higgsing the theory obtained from the orbifold projection.
It is thus illustrative for us to present an analysis for these parent orbifold theories. The toric diagram is an $k \times k$ right isosceles triangle of lattice points and the matter content is captured by a periodic quiver with $k^2$ nodes and the superpotential is comprised of the closed triangles in the quiver (cf.~\cite{Hanany:1997tb} which obtained the first results for these $\cN=1$ gauge theories):
\begin{equation}
\includegraphics[trim=0mm 0mm 0mm 0mm, clip, width=4in]{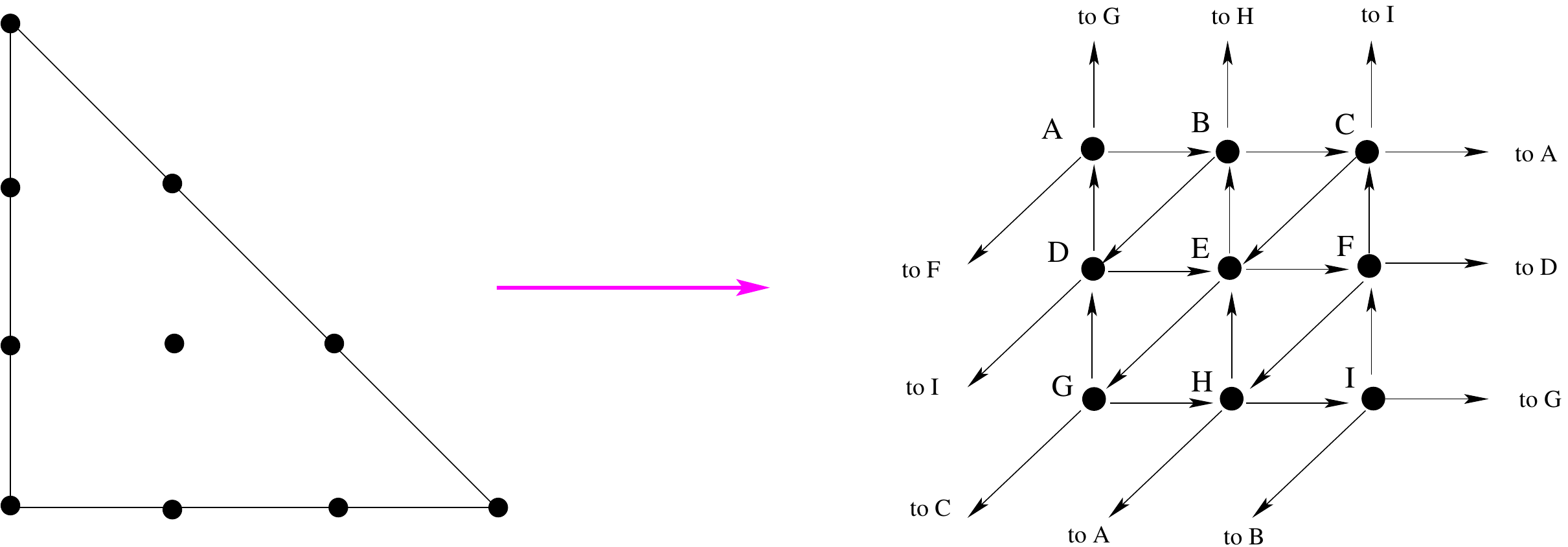}
\end{equation}
The plethystic programme was carried out for these spaces in \cite{Feng:2007ur} and the VMS was found to be complete intersection; in particular, the Hilbert series and syzygies are presented in Eqs (4.4-4.5) in {\it cit.~ibid.}:
\begin{equation}
g_1(t; \IC^3/\IZ_k^2) = f_\infty(t; \IC^3/\IZ_k^2) = 
   \frac{1-t^{3k}}{(1-t^3)(1-t^k)^3} \ . 
\end{equation}

At $k=1$, we are of course back to the case of $\IC^3$.
For $k=2$, we have the cone over a single sextic in $W\IP^3_{[2:2:2:3]}$ from the above expression for $g_1$. Writing the c\"oordinates of the weighted projective space (in the order of the prescribed weights) as $x,y,z,w$, we have the sextic as $xyz=w^2$. Hence,
\begin{eqnarray}
\nn
g_1(t; \IC^3/\IZ_2^2) &=& f_\infty(t; \IC^3/\IZ_2^2) = 
   \frac{1-t^{6}}{(1-t^3)(1-t^2)^3} = 
\sum\limits_{n=0}^\infty {n + \frac52 + \frac32(-1)^n \choose 2}t^n \ , \\
g_\nu(t; \IC^3/\IZ_2^2) &=&
   \prod\limits_{m=0}^\infty 
   (1 - \nu t^m)^{-{n + \frac52 + \frac32(-1)^n \choose 2}}
 = 1 + \frac{1-t^{6}}{(1-t^3)(1-t^2)^3} \nu + \\
\nn
&& + \frac{3 t^6-5 t^5+4 t^4-t^3+2 t^2-2 t+1}{(1-t)^6 (1+t)^4
   (1+t^2)^2} \nu^2 + \cO(\nu^3) \ .
\end{eqnarray}
For the weighted projective variety we find that $b_{0,\ldots, 4} = \{1,0,0,0,1\}$. Hence the zeta function is actually quite simple, because the base space is homologically rather trivial.
The counting for the affine cone thus proceeds as though it were $\IC^3$ and we have that
\begin{equation}
Z_p(t; \IC^3/\IZ_2^2) = (1-p^3t)^{-1} \ , \quad
Z(s; \IC^3/\IZ_2^2) = \zeta(s-3) \ .
\end{equation}
Indeed, explicitly counting over the first 20 primes on the computer confirms the $p^3$ solution.
We remark that had we desingularized the origin to $xyz-w^2 = \epsilon$ and set $\epsilon = 1$ to avoid primes of bad reduction, the solutions are drastically different:
\[
N_{p^{r=1}} = \{8, 12, 170, 252, 1100, 2522, 5474, 6156, 11132, 26042,27900, 53354, 72242,\ldots\}
\]

As our last remark, the (irreducible top-dimensional component of the) master space of this example was studied in detail in \cite{Forcella:2008bb} and we recall, from Eq (2.11) therein, that it is a Calabi-Yau variety of dimension 6, degree 14 and comprised of the incomplete intersection of 15rics in 12 variables, with the Hilbert is given by $f(t; \firr{\IC^3/\IZ_2^2}) = (1+6t+6t^2+t^3)(1-t)^{-6}$.
Reducing over the first primes we obtain $N_{p^{r=1}} = \{136, 1377, 24625, 167041, 2250721 \ldots \}$.

The next simplest case is the hypersurface $xyz = w^3$ corresponding to $k=3$ (which partially resolves to all the toric del Pezzo cones, to which we shall shortly return). 
This is an affine cone over the cubic in $\IP^3$ and we can rescale weights to obtain
\begin{eqnarray}
\nn
g_1(t; \IC^3/\IZ_3^2) &=& f_\infty(t; \IC^3/\IZ_3^2) = 
   \frac{1-t^{3}}{(1-t)^4} = 
\sum\limits_{n=0}^\infty \left( \frac{3n(n-1)}{2} + 1 \right)t^n \ , \\
g_\nu(t; \IC^3/\IZ_3^2) &=&
   \prod\limits_{m=0}^\infty (1 - \nu t^m)^{-\frac{3n(n-1)}{2} - 1} \\
\nn
&=& 1 + \frac{1-t^{3}}{(1-t)^4} \nu +
+ \frac{4 t^6+4 t^5+11 t^4+7 t^3+8 t^2+t+1}{(1-t)^3
   (1-t^2)^3} \nu^2 + \cO(\nu^3) \ .
\end{eqnarray}
Now, the cubic surface in $\IP^3$ is the well-known del Pezzo surface which is $\IP^2$ blown up at 6 generic points and whose Betti numbers are standard: $b_{0,\ldots, 4} = \{1,0,7,0,1\}$. This is confirmed by the toric diagram, which captures a special point in the moduli space of this del Pezzo surface (after all, only up three blow-ups can be accommodated by a toric description).
Therefore, the zeta function has the form, following the argument above for the conifold, {\it mutatis mutandis}:
\begin{equation}
Z_p(t; \IC^3/\IZ_3^2) = 
\frac{1}{(1-t)^2} \frac{1}{(1-p^2 t)}\frac{1}{P^{(7)}(t)} \ ,
\end{equation}
where $P^{(7)}(t)$ is a polynomial of degree 7 with integer coefficients and unit constant term.
We may then use the values of $N_{p^i}$ for $i=1, \ldots, 6$ to fix the indeterminate coefficients in $P^{(7)}(t)$ much like what we did for the conifold, an intense computation into which we shall not presently delve.
Suffice it to say that, trying out the first number of primes gives us $N_{p^{r=1}} = p^3$ and the first few values of $N_{p^{r=2}}$ are $96, 1107, 22125, 156751\ldots$

Indeed, as $k$ grows, the degree of the indeterminate factors too will grow, and whence the number of coefficients to fix.
For now, we shall not occupy ourselves with the higher cases, though it is certainly interesting to find out what the growth rates of the coefficients in the zeta function are with respect to $k$.

\subsection{Del Pezzo Cones}
The importance of del Pezzo surfaces, in their ubiquitous appearances in algebraic geometry, representation theory as well as gauge theory, can hardly be over-stated.
In the framework of D3-brane probes, they provide a marvellously rich class of stringy background by being the base surfaces over which cones are affine Calabi-Yau threefolds.
The world-volume gauge theories for the toric members ({\it viz.} $\IP^2$ blown up at $n=0,\ldots,3$ points, as well as $\IP^1 \times \IP^1$) were first presented in \cite{Feng:2000mi} while the higher ones ($n=4,\ldots8$) were given in \cite{Wijnholt:2002qz}.
It is irresistible that we at least mention these gauge theories.

In \cite{Benvenuti:2006qr}, we found that for the cone over the $m$-th del Pezzo surface (and hence of degree $9-m$), the fundamental generating function is
\begin{equation}
f(t; dP_m) = \frac{1+(7-m)+t^2}{(1-t)^3} \ , \qquad m = 0, \ldots, 8 \ .
\end{equation}
The case of $m=6$, {\it i.e.}, the cubic surface, we have already probed in our aforementioned study of the $\IZ_3 \times \IZ_3$ orbifold, as a special point in the complex structure moduli space, so here let us move onto another simple example, say, $m=0$. 
This is simply the total space of the $\cO(-3)$ anti-canonical bundle over $\IP^2$, resolving the Gorenstein singularity $\IC^3 / \IZ_3$.
As an affine embedding, this is given by the non-complete intersection of 27 quadrics in $\IC^{10}$, explicitly presented in Eq (5.14) of \cite{Benvenuti:2006qr}.

Because the Betti numbers of the base are $b^{i=0,\ldots,4} = (1,0,1,0,1)$ and that we are adding the origin as the tip of the cone, we have, as before, that
$Z_p(t; dP_0) = \frac{1}{(1-t)^2(1-pt)(1-p^2t)}$,
being fixed by \eqref{Zp} after we projectivize back to $\IP^2$.
Therefore, the global zeta function is $Z(s;dP_0) = \zeta(s)^2\zeta(s-1)\zeta(s-2)$.

As a first non-trivial example, let us consider the next del Pezzo cone, of $m=1$. Here, we have $f(t; dP_1) = (1-t)^{-3}(1+6+t^2)$ and
\begin{equation}
PE^{-1}[f(t; dP_1)] = 9 t - 20 t^2 + 64 t^3 - 280 t^4 + 1344 t5 + \cO(t^6) \ ,
\end{equation}
signifying a non-complete intersection.
The gauge theory can be found in Section 4 of \cite{Feng:2000mi}, with the adjacency matrix $a_{ij}$ of the quiver and the superpotential $W$ given by
\begin{equation}
a_{ij} = {\scriptsize \left(\begin{matrix}
0 & 1 & 1 & 0 \\
 0 & 0 & 2 & 0 \\
 0 & 0 & 0 & 3 \\
 2 & 1 & 0 & 0
\end{matrix}\right)} \ ,  \qquad
\begin{array}{rcl}
W &=& -X_{4,2} X_{2,3}^2 X_{3,4}^1+X_{1,3} X_{4,1}^2 X_{3,4}^1+X_{4,2}
X_{2,3}^1 X_{3,4}^2- \\
&& -X_{1,3} X_{3,4}^2 X_{4,1}^1+X_{1,2} X_{2,3}^2 X_{3,4}^3
X_{4,1}^1-X_{1,2} X_{2,3}^1 X_{3,4}^3 X_{4,1}^2 
\end{array}
\ ,
\end{equation}
where we have used the standard notation that $X_{i,j}^k$ is the $k$-th arrow from nodes $i$ to $j$.
The VMS is readily found by the methods outlined in \S\ref{s:vms} and comprises of 20 quadrics in $\IC^9$:
\begin{equation}
\begin{array}{rcl}
VMS(dP_1) &\simeq& \{
y_{3}^2-y_{1} y_{4},y_{2} y_{3}-y_{1} y_{5},y_{2} y_{5}-y_{1} 
y_{6},y_{2}y_{4}-y_{1} y_{7},y_{3} y_{5}-y_{1} y_{7},y_{4} y_{5}-y_{3} 
y_{7},\\
&&
y_{5}^2-y_{1}y_{8},y_{3} y_{6}-y_{1} y_{8},y_{2} y_{7}-y_{1} y_{8},y_{4} 
y_{6}-y_{3} y_{8},y_{5}
   y_{7}-y_{3} y_{8},y_{7}^2-y_{4} y_{8},\\
&&
y_{5} y_{6}-y_{1} y_{9},y_{2} y_{8}-y_{1}
   y_{9},y_{6}^2-y_{2} y_{9},y_{6} y_{7}-y_{3} y_{9},y_{5} 
y_{8}-y_{3} y_{9},y_{7}
   y_{8}-y_{4} y_{9},\\
&&y_{6} y_{8}-y_{5} y_{9},y_{8}^2-y_{7} y_{9}
\} \ .
\end{array}
\end{equation}
One can also check \cite{m2} that the Hilbert series for this embedding is as stated.

Now, the base surface is a projective variety in $\IP^8$ and has Betti numbers $b^{i=0,\ldots,4} = (1,0,2,0,1)$, thus the local zeta function should be
\begin{equation}
Z_p(t; dP_1) = \frac{1}{(1-t)^2(1-p^2t)[(1-\alpha t)(1-\beta t)]}
\end{equation}
where $\alpha$ and $\beta$ are constants to be determined.
Proceeding as in \eqref{NpAp}, we in fact find the same number of zeros as \eqref{Npconi} and whence the same form of the local and global zeta functions as the conifold.

\subsection{SQCD}
Having indulged ourselves with a plethora of examples, many from the context of D-branes and Calabi-Yau spaces in string theory, let us, as our final set of illustrations, take a complete departure and study perhaps the most canonical field theory of them all, {\it viz.}, supersymmetric QCD.
The geometry of this was the theme of \cite{Gray:2008yu}, whose intent was to provide an algebro-geometric and plethystic aper\c{c}u on this old subject.
The explicit VMS as an affine variety was computed and some first examples, presented in Eq (3.25) therein.

Let us denote the VMS of SQCD with $N_f$ flavours and $N_c$ colours as $\cM_{(N_f,N_c)}$. Then for $N_f < N_c$, $\cM_{(N_f,N_c)} \simeq \IC^{N_f^2}$, the flat affine space. For $N_f \ge N_c$, $\cM_{(N_f,N_c)}$ has complex dimension $2N_cN_f - (N_c^2-1)$ as an affine variety embedded in $\IC^{N_f^2+2}$. In the particular case when $N_f=N_c$, $\cM_{(N_c,N_c)}$ is a single hypersurface of degree $2N_c$.
For some first few values, the defining equations of the VMS and the associated Hilbert series are:
\begin{equation}
\begin{array}{|l|l|l|}\hline
M_{1,1} & \{-y_1 + y_2 y_3 \} & \frac{1}{(1-t)^{2}}\\ \hline
M_{2,1} & \{-y_{6}y_{8}+y_{4},-y_{5}y_{8}+y_{2},-y_{6}y_{7}+y_{3},
            -y_{5}y_{7}+y_{1}\} & \frac{1+t4+t^2}{(1-t)^4}\\ \hline
M_{2,2} & \{ y_{2}y_{3}-y_{1}y_{4}+y_{5}y_{6} \} & 
  \frac{1+t^2}{(1-t^2)^5} \\ \hline
M_{3,3} & \{y_{3}y_{5}y_{7}-y_{2}y_{6}y_{7}-y_{3}y_{4}y_{8}
   +y_{1}y_{6}y_{8}+y_{2}y_{4}y_{9}-y_{1}y_{5}y_{9}+y_{15}y_{21} \}
& \frac{1+t^3}{(1-t^2)^9(1-t^3)}
\\
\hline
\end{array}
\end{equation}

Now, $\cM_{(1,1)}$ is just the flat space $\IC$ and needs no further comment.
$\cM_{(2,1)}$ is a dimension 4, degree 6 affine variety. Counting the number of points, reducing over the first few primes, shows that here $N_{p^r} = (p^4)^r$.
Going with this pattern easily gives us that
\begin{equation}
Z_p(t; \cM_{(2,1)}) = \frac{1}{1-p^4t} \ , \quad
Z(s; \cM_{(2,1)}) = \zeta(s-4) \ ,
\end{equation}
as though we have the counting for $\IC^4$.
$\cM_{(2,2)}$ is more complicated; this is a degree 2, dimension 5 variety as a cone over a quadric 4-fold. The Betti numbers are easily determined to be
$\{b_{0,\ldots,8}\} = \{1,0,1,0,2,0,1,0,1\}$, for an Euler number of 6.
Therefore, the local zeta function, by \eqref{Zp} and adding the tip of the affine cone, is
\begin{equation}
Z_p(t; \cM_{(2,2)}) = \frac{1}{(1-t)^2 (1-pt)(1-B_pt+pt^2)(1-p^3t)(1-p^4t)}
\ .
\end{equation}
\comment{
where the extra factor of $(1-t)^{-1}$, as above, is count the point at the origin as we include the affine cone.}
We have indeterminate coefficients $B_p$ which we can fix by observing some values. 
As was in the case of the conifold, we expand the above expression to relate the coefficient $B_p$ with $N_{p^{r=1}}$, the latter of which we can list the first few values:
\begin{equation}
N_{p^{r=1}} = 2 + p + p^3 + p^4 + B_p = 
\{ 36, 261, 3225, 17101,162261, 373321, 1424481 \ldots\}
\end{equation}
Hence, the global Hasse-Weil zeta function becomes
\begin{equation}
Z(s; \cM_{(2,2)}) = \zeta(s)^2\zeta(s-1)\zeta(s-3)\zeta(s-4)L(s; \cM_{(2,2)}) \ ,
\end{equation}
where the L-function is $L(s; \cM_{(2,2)}) = 
\prod\limits_p (1-B_pp^{-s}+p^{1-2s})^{-1}$ with the first few values of $B_p$ being $\{8, 148, 2468, 14348, 146276, 342548, 1336028  \ldots \}$.
Developing $Z(s; \cM_{(2,2)})$ into a Dirichlet series gives the first few coefficients as $c_n = \{1, 36, 261, 841, 3225, \ldots \}$.

\section{Generationes et Generationes}\label{s:compare}\setall
We have thus performed extensive experimentation, in studying the gauge invariant as well as the arithmetic properties of a host of supersymmetric field theories. The data presented are perhaps of interest {\it ipso facto}. However, contented with our catalogue, we could forge ahead with some further calculations.
That two enumeration problems, as seen from the proceeding discussions, should each be governed by a rational function as a generating function naturally lends itself to an instant speculation.
Could the zeta function of the VMS of one gauge theory, encoding its zeros over finite fields, be related to the Hilbert series of the VMS of another gauge theory?
This comparative study, relating one generation to another \footnote{Hence the title of the section: {\it corpora ipsorum in pace sepulta sunt et nomen eorum vivet in generationes et generationes. - Ecclesiasticus 44:14}.},
would engender quite a peculiar relationship, wherein the BPS mesonic spectrum of a gauge theory should correlate to the arithmetic of another vacuum geometry.

Two immediate hurdles, however, quickly present themselves were we to make a na\"{\i}ve identification.
First, the (local) zeta function, which is a rational function according to Weil-Deligne, is defined with respect to a given prime number $p$; the straight-forward analogue of this parametre in the case of the Hilbert series is unclear.
It seems unnatural that the BPS spectrum should be at all particular to any fixed prime number.
Second, and perhaps more seriously, is the difference in the growth rate of $a_n$, the number of gauge invariant operators versus that of $N_{p^n}$, the number of zeros over $\IF_{p^n}$.
The former, being the Hilbert series of an algebraic variety, usually tends polynomially in $n$ (and indeed, governed by the so-called Hilbert polynomial in the degree $n$). 
The latter, however, grows rather much faster, and is in fact exponential in $n$, say $\sim p^n$.
This is not only seen in the examples presented in \S\ref{s:zeta}, but is, in fact, compelled to be thus by \eqref{Zp}, so as to ensure that the generating zeta function can have the exponent behaving logarithmically, and consequently cancelling the exponential to give a rational function.

For example, $a_n \sim n$ is a perfectly acceptable growth for a gauge theory.
In fact, the mesonic spectrum of the D-brane theory on the flat-space $\IC^2$, or equivalently, the Hilbert series of the bi-variate polynomial ring $\IC[x,y]$, is simply $f_\infty(t; \IC^2) = (1-t)^{-2}$; whence $a_n = n+1$.
However, having $N_{p^n} = n+1$ would force the zeta function to be non-rational; signifying that no algebraic variety over any number field could possibly have such a behaviour for its zeros.

Parenthetically, we point out that in an interesting paper \cite{HomoAlg}, the authors find a fascinating relation between the Hilbert series of a variety and the zeta function of another \cite{HSzeta}.
There, the Veronese curve $X$ prescribed by the embedding of $\IP^1$ by the very ample line bundle $L = \cO_{\IP^1}(P+1)$ for some $P \in \IZ_+$ is considered.
The dimension at degree $n$ is therefore $h^0(\IP^1, \cO_{\IP^1}(P+1))$, giving us the Hilbert series:
\begin{equation}
H(t;X) = \sum_{n=0}^\infty ((P+1)n+1)t^n = \frac{1 + P t}{(1-t)^2} \ .
\end{equation}
On the other hand, we recall that the zeta function for $\CA^1$, given momentarily in the discussion on that of $\IP^k$, is $(1-pt)^{-1}$.
We thus see that the numerator of the Hilbert series of the Veronese curve, evaluated at the negative of its argument, identifies with the denominator of the Weil zeta function of the affine line, a seemingly different geometry.
In this example, the aforementioned first objection was circumvented by the choice of the line bundle in embedding $\IP^1$, whereby inherently introducing a parametre $P$, which is then associated with some prime $p$.

Similarly, one could consider the $\nu$-inserted plethytics for $\IC^1$.
Here, as was computed in \cite{Benvenuti:2006qr}, $H(t; \IC^1) = (1-t)^{-1}$, giving $a_n=1$, and whence $g(\nu, t ; \IC^1) = \prod\limits_{m=0}^{\infty} (1 - \nu t^m)^{-1} = 1 + \sum\limits_{N=1}^{\infty} g_N(t; \IC^1) \nu^N$, with $g_N(t; \IC^1) = \prod\limits_{n=1}^N (1 - t^n)^{-1}$.
This, with a simple re-definition, is of the form for the zeta function for $\IP^k$ in \eqref{zetaPN}.

These above digressions are, of course, merely formal resemblances.
What we wish for is a systematic correspondence; this is certainly encouraged by the similarity between the definitions of the zeta function and the plethystic exponential, a similarity whose discrepancies, however, are of sufficient significance that a direct identification is not pronouncedly manifest.
Nevertheless we are inspired by the following diagram:
\begin{equation}\label{diag}
\hspace{-1cm}
\begin{array}{|c|c|c|c|c|}\hline
\mbox{\begin{tabular}{l}Defining\\ Quantities\end{tabular}}
&
&
\mbox{\begin{tabular}{l}Rational\\ Function\end{tabular}}
&
&
\mbox{\begin{tabular}{l}Global\\ Objects\end{tabular}}
\\ \hline\hline
\mbox{
\begin{tabular}{l}
Syzygies $s(t)$\\
(Geometric)\\
\end{tabular}
}
&
\stackrel{\exp{\sum\limits_{r=1}^\infty\frac{s(t^r)-s(0)}{r}}}{\xrightarrow{\hspace*{2cm}}}
&
\mbox{\begin{tabular}{l}Hilbert Series \\ $f(t) = \sum\limits_{n=0}^\infty a_n t^n$\end{tabular}}
&
\stackrel{\exp{\sum\limits_{r=1}^\infty\frac{f(t^r)-f(0)}{r}}=\prod\limits_{n=1}^\infty (1-t^n)^{-a_n}=\sum\limits_{n=0}^\infty d_n t^n}{\xrightarrow{\hspace*{5cm}}}
&
\mbox{Full Spectrum}
\\ \hline
\mbox{
\begin{tabular}{c}
Rational\\ 
Points $N_{p^r}$\\
(Arithmetic)\\
\end{tabular}
}
&
\stackrel{\exp{\sum\limits_{r=1}^\infty\frac{N_{p^r}}{r}t^r}}{\xrightarrow{\hspace*{2cm}}}
&
\mbox{\begin{tabular}{l}Zeta Function \\$Z_p(t=p^{-s})$\end{tabular}}
&
\stackrel{\prod\limits_p Z_p = Z(s) = \sum\limits_{n=1}^\infty \frac{c_n}{n^s}}{\xrightarrow{\hspace*{3cm}}}
&
\mbox{Dirichlet Series}
\\ \hline
\end{array}
\end{equation}

The diagram is self-suggestive and let us make a few remarks.
The geometric object of concern is $X$, the classical VMS of a gauge theory.
The syzygies $s(t)$, or defining equations of $X$, is a polynomial in $t$ when $X$ is complete intersection, otherwise, it will be some power series.
The plethystic exponential takes $s(t)$ to the Hilbert series $f(t)$, which counts the single-trace (mesonic) BPS spectrum of the gauge theory and is a rational function by Hilbert's theorem on algebraic varieties. The full (mesonic) spectrum is obtained by a second plethystic exponentiation, in the spirit of a bosonic oscillator partition function and recast as a product over the vibration modes, from $f(t)$. 
These multi-trace operators constitute the ``global'', or complete, set of objects in the gauge theory, and is obtainable from two plethystic substitutions from the intrinsic geometric property $s(t)$ of $X$.

In a parallel vein, we can consider the arithmetic properties of $X$ and enumerate the number of solutions $N_{p^r}$ over finite fields. An exponential generating function gives the zeta function, which is rational by \eqref{Zp}. 
This is made global by a product over primes and gives a Hasse-Weil zeta function, which can then be expanded into a Dirichlet series.
The plethystics have analytic inverses involving the M\"obius function while the inverse procedure to forming the zeta functions is quite difficult.
Nevertheless, the similarities of proceeding from intrinsic geometric (or arithmetic) properties of $X$, via a rational function, to a global enumerative problem associated with the gauge theory, through two exponential (infinite product) substitutions in generating functions, is tantalizing indeed.

\subsection{From Hasse-Weil to Hilbert and Back}
Let us commence again with examples.
Since the zeta function is severely restricted in form, it is perhaps expedient to start therewith.
So our strategy will be to begin with a gauge theory whose VMS is $X$, we then compute the arithmetical properties of $X$, starting from the bottom left box in the diagram in \eqref{diag}, trace the arrows, via rationality and globality, to the right and then go upwards, trace the arrows backwards, via locality and syzygy, to the geometrical properties of a possibly different VMS $Y$ of another gauge theory.

\subsubsection{A Single Point}
Let $X$ be a single point.
We recall from \eqref{euler} that here the local zeta function is $(1-t)^{-1}$ and the global zeta function is the Riemann zeta function. The L-series coefficients $c_n$ are thus all unity. We then identify $c_n$ with the coefficients $a_n$ in the plethystic exponential, and can therefore form the power series $1 + \sum\limits_{n=1}^\infty 1 \ t^n = (1-t)^{-1}$, where we have added 1 as the zeroth term for normalization in order to take care of the $f(0)=1$ term in the plethystic exponent.
We have arrived, of course, at the Hilbert series for $\IC$, and subsequently the mesonic BPS operators of D-branes probing this trivial Calabi-Yau 1-fold.
The gauge theory corresponding to this VMS, as we recall from \S\ref{s:1field},
is a free theory of a single field.
\comment{
It consists of a single adjoint field $\Phi$ charged under $U(N)$ at large $N$ and can be thought of $\cN=4$ in 4-dimensions.
The BPS invariant operators are simply $\tr \Phi^k$ for $k=1,2,3,\ldots$ and the VMS is simply the complex line $\IC$ wherein $\tr \Phi$ can take value.
At fixed finite $N$, the $\nu$-insert plethystic gives that \cite{Benvenuti:2006qr} $f_N(t; \IC) = \frac{1 - t^{N+1}}{1-t}$ and that $g_N(t; \IC) = \prod\limits_{n=1}^N (1-t^n)^{-1}$. The VMS for give $N$ is the $N$-th symmetric product of $\IC$, {\it i.e.}, $\IC^N / \Sigma_N$ (with $\Sigma_N$ the symmetric group on $N$ elements, permuting the $N$-copies of $N$) the Hilbert Series for which is prescribed by $f_N(t; \IC)$.
}
Thus, in our trivial warm-up example, we have gone from a point to $\IC$, which can be thought of as a cone over a point.

\subsubsection{Affine Space}\label{s:cn}
Next, let us study the family of affine space $\IC^k$; this can be thought of as the VMS of $k$ mutually commuting $SU(N)$ fields at large $N$.
From \eqref{zetaCN}, we recall that $Z(s; \IC^k) = \zeta(s-k) = \sum\limits_{n=1}{n^{s-k}}$.
Therefore the Dirichlet coefficients are $c(n) = n^k$.
Were this the enumerations of a Hilbert series, we would have that (again normalizing with 1 for the $n=0$ term):
\begin{equation}
F(t; \IC^k) = 1 + \sum\limits_{n=1}^\infty n^k t^n = 1 + \li_{-k}(t) \ ,
\end{equation}
where $\li$ is the standard (de Jonqui\`ere's) Poly-Logarithm function.
Note the nomenclature here: we have used $F(t; \mbox{space})$ because the convention in the preceding discussion was that $f(t; X)$ refers to the Hilbert series of the algebraic variety $X$ whereas here we are formally constructing a power-series, the space for which $F(t)$ may be a Hilbert series is yet to be determined.

Now, for integral parametres $k$, $\li_{-k}(t)$ are all rational functions, which is re-assuring as the corresponding Hilbert series should be so as is required by an algebraic variety.
Specifically, let us recall some of the first few values:
\begin{eqnarray}
\nn \li_{0}(t) &=& \frac{t}{1-t} \ , \quad
\li_{-1}(t) = \frac{t}{(1-t)^2} \ , \quad
\li_{-2}(t) = \frac{t (1+t)}{(1-t)^3} \ , \\
\li_{-3}(t) &=& \frac{t \left(1 + 4 t + t^2\right)}{(1-t)^4} \ , \quad
\li_{-4}(t) = \frac{t \left(1 + 11 t + 11t^2 + t^3\right)}{(1-t)^5} \ .
\end{eqnarray}
Indeed, for $k=0$, this gives us $f(t) = (1-t)^{-1}$, in accord with the aforementioned case of the single point.
Subsequently, we have that:
\begin{eqnarray}
\nn 
F(t; {\rm pt}) &=& \frac{1}{1-t} \ , \\
\nn
F(t; \IC) &=& \frac{1-t+t^2}{(1-t)^2} \ , \\
\label{FCN}
F(t; \IC^2) &=& \frac{1 - 2t + 4t^2 -t^3}{(1-t)^3} \ , \\
\nn
F(t; \IC^3) &=& \frac{1 - 3t + 10t^2 - 3t^3 +t^4}{(1-t)^4} \ , \\
\nn
F(t; \IC^4) &=& \frac{1 - 4t + 21 t^2 + t^3 - 6t^4 - t^5}{(1-t)^5} \ .
\end{eqnarray}
That these functions are in the form of Hilbert series of either the first or second kind is pleasing.

Note that the numerators for $\li$ are all palindromic; for a Hilbert series, this would imply that the corresponding algebraic variety be Calabi-Yau by Stanley's theorem \cite{stanley} (cf.~\cite{Forcella:2008bb} for its implication in D-brane gauge theories). However, we see that the full Hilbert series is palindromic only for odd $k$. 
This can be seen from the following argument.
First, we have the so-called inverse formula \cite{GR}, that for all $s \in \IC$,
\begin{equation}
\li_s(z) + (-1)^s \li_s(1/z) = \frac{(2\pi i)}{\Gamma(s)} \times \left\{
\begin{array}{ll}
\zeta(1-s, \frac12 + \frac{\log(-z)}{2 \pi i}) \ , & z \notin (0,1] \\
\zeta(1-s, \frac12 - \frac{\log(-1/z)}{2 \pi i}) \ , & z \notin (1,\infty)
\end{array}
\right. \ ,
\end{equation}
where $\zeta(a,z) := \sum\limits_{n=0}^\infty (a+n)^{-s}$ is the standard Hurwitz zeta function.
Because of the pole of the Gamma function at negative integer values, this implies that
\begin{equation}\label{li-palin}
\li_{-n}(z) + (-1)^n \li_{-n}(1/z) = 0 \ , \qquad n = 1,2,3, \ldots
\end{equation}
Now, because each of our Polylogarithmic function with negative integral parametre is a rational function with equal degree of numerator and denominator, with the latter being trivially $(1-t)^{k+1}$, palindromicity of the numerator simply means that $F(t)$ should equal $F(1/t)$, which is indeed guaranteed by \eqref{li-palin} for odd $k$.

Returning to our list in \eqref{FCN}, the first case of $k=0$ is simply the point.
Next, with the case of $k=1$ and hence $F(t; \IC)$ we also have some familiarity.
Let us take the plethystic logarithm to yield
\begin{equation}\label{T2}
PE^{-1}[\frac{1-t+t^2}{(1-t)^2}] = t + t^2 + t^3 - t^6 \ .
\end{equation}
According to the rules prescribed in \cite{Benvenuti:2006qr} and some discussions on a similar circumstance in \cite{Feng:2007ur}, if we were to construe the above as a projective variety, then the terminating plethystic logarithm, signifying the syzygies, should be interpreted as follows:
we have three generators, in degrees 1,2 and 3 respectively, obeying a single relation in degree 6. 
That is, we have a sextic hypersurface in weighted projective space $W\IP^2_{[1:2:3]}$.
This, of course, is none other than an elliptic curve.

In light of our present discussion, that we are dealing with affine spaces and that our gauge theory VMSs are naturally {\it affine} varieties, it is perhaps more expedient of interpret the Hilbert Series as that of an affine variety.
Thus, we think of the geometry above as that of an affine cone over ({\it i.e.}, the dehomogenization of) the said elliptic curve.

The gauge theory is not immediately reconstructible.
Indeed, were the moduli space simply $\IC^2$, then we would have, much in analogy with \S\ref{s:1field}, an $SU(N)$ theory with two adjoint fields at large $N$, say $\Phi_1$ and $\Phi_2$, such that the F-terms from the superpotential force them to commute and all the mesonic BPS operators would be in the form $\tr(\Phi_1^{n_1}\Phi_2^{n_2})$ with $n_1,n_2 \in \IZ_{\ge 0}$. However, the Hilbert series would simply be $(1-t)^{-2}$, with trivial numerator.
\comment{NOT QUITE right: this gives one without the numerator and in fact gives g_3 of C^3.
Nevertheless, we have some familiarity  (cf.\cite{Hanany:2008gx}) with the affine variety associated to $F(t;\IC)$: this is simply the quotient singularity $\IC^3 / \Sigma_3$, where if we assign $x,y,z$ as the co\"ordinates of $\IC^3$, the symmetric polynomials of degrees 1, 2 and 3 are precisely the primitive invariants required, having a syzygy at degree 6.
Therefore, we can think of the gauge theory as that being prescribed by a D3-brane probing the Gorenstein quotient singularity $\IC^3 / \Sigma_3$, which is also the dihedral group embedded in $SU(3)$ \cite{Hanany:1998sd}, giving a 4-dimensional $\cN=1$ quiver gauge theory.}
Nevertheless, we see that $F(t; \IC)$ is written in the form of a Hilbert series of second kind. Thus, the associated affine variety is of complex dimension 2 and of degree $P(1) = 1$ where $P(t) = 1-t+t^2$ is the numerator.
This can geometrically be realized as a line in $\IC^3$. 
Interestingly, the multi-trace operators are counted by
\begin{equation}\label{macmahon}
PE[\frac{1-t+t^2}{(1-t)^2}] = \prod\limits_{n=1}^\infty \frac{1}{(1-t^n)^n} :=
Mac(t) \ ,
\end{equation}
the MacMahon function, which is crucial to the crystal melting picture of A-type topological strings \cite{Okounkov:2003sp}.
It is curious that so rich a structure can be encoded in the zeta function of so simple a geometry as the affine complex line.

Moving on to higher $k$ gives more involved results: indeed, one can check that the plethystic logarithm of $F$ in \eqref{FCN} are {\it non-terminating}.
This means that the corresponding algebraic varieties are not obviously complete intersections and there exist non-trivial higher syzygies {\it ad infinitum}.
Again, we can interpret the Hilbert series as being of the second kind since the numerator and denominator have been cleared maximally.
Hence, $k$ corresponds to a variety of dimension $k+1$ and degree $k!$, following readily from the asymptotic expansion of the poly-logarithm, that
\begin{equation}
\li_{-k}(e^\mu) = \frac{k!}{(-\mu)^{k+1}} - \sum\limits_{m=0}^\infty
  \frac{B_{k+m+1}\mu^m}{m! (k+m+1)} \ ,
\end{equation}
where $B$ are the Bernoulli numbers and the limit $\mu \to 0$ should be taken.

Let us focus, as inspired by string theory, on cases of $k=2,3$, these being potentially theories on the D3 and M2-branes \cite{Hanany:1998sd,Hanany:2008gx}.
Taking the plethystic logarithm of the $k=2$ case gives
\begin{equation}
PE^{-1}[F(t;\IC^2)] = 
t+3 t^2+5 t^3+t^4-6 t^5-17 t^6-4 t^7+29 t^8+56 t^9+7 t^{10}
+\cO(t^{11}) \ .
\end{equation}
Indeed, as much as $F(t; \IC)$, whose expansion coefficients are $n$, needs to be compared to $\IC^2$, whose coefficients are $n+1$, the present example, whose expansion coefficients are $n^2$, needs to be compared to the {\it conifold} (quadric hypersurface defined in $\IC^4$), which has coefficients $(n+1)^2$ as was seen in \eqref{fCon}.
Both latter cases are simple geometries whereas the former, by a seemingly trivial shift of 1, already becomes quite involved.
Moreover, since the numerator is not palindromic, the space needs not even be Calabi-Yau.
In any event, we know this to be a 3-dimensional variety of degree 2.
Similarly, for $k=3$, we have that
\begin{equation}
PE^{-1}[F(t;\IC^3)] = 
t+7 t^2+19 t^3+9 t^4-72 t^5-246 t^6-72 t^7+1422 t^8+\cO\left(t^{9}\right) \ ,
\end{equation}
a non-complete intersection space of dimension 4 and degree 6.
Therefore, in each case of $k$, we could indeed find a VMS of a rather non-trivial gauge theory whose Hilbert series coincides with the zeta function of $\IC^k$.

\subsubsection{Conifold Revisited}
The astute reader may have asked why, in the above example, in identifying the Dirichlet coefficients from the Euler product with plethystic product, we have set $c_n$ equal to $a_n$, rather than $d_n$ (adhering to the nomenclature of \eqref{diag}).
The reason, of course, is that we wish to readily guarantee that we could arrive at an obvious rational function upon taking the plethystic logarithm.
Indeed, as we shall discuss in further detail in \S\ref{s:asym}, we need to be mindful of the growth rate of these coefficients.
Begotten from a Hilbert series of commutative variety, $a_n$ are approximately polynomial growth, and whence, by plethystic exponentiation, $d_n$ grow as a polynomial multiplied by an exponential.

On the other hand, due to the work of Schnee, Landau and Ramanujan on the general theories of Dirichlet series (cf.~\cite{HR}), $c_n$ is usually taken to be of polynomial growth in order to allow absolute convergence of the Dirichlet series in the upper half-plane and its subsequent analytic continuation - crucial, of course, for any statements pertaining to the generalized Riemann Hypothesis.
Therefore, identification of $c_n$ with $a_n$ is the more natural choice for now, as was seen above.
Indeed, for algebraic schemes more miscellaneous, the Hilbert series can be arbitrary and whence more general identifications could be permissible.

Let us illustrate with the example of the conifold on which we expounded to some length above.
We recall the Dirichlet coefficients $c_n$ from \eqref{DirichletC} and redevelop this as a power series (without troubling ourselves too much with convergence presently).
Setting these to be the $a_n$ coefficients in a power series, we can then take the plethystic logarithm, at least order by order, to find that
\begin{eqnarray}
\nn
f = PE^{-1}[Z(s;\widetilde{\cC})] &=& 
7 t-3 t^2-31 t^3+209 t^4-744 t^5+1431 t^6+2194 t^7-35726 t^8+\\
&&+186120t^9-573070 t^{10}+\cO\left(t^{11}\right)
\end{eqnarray}
According to the rules, this would describe a non-complete-intersection manifold, generated by 7 linear forms, obeying 3 quadratic, 31 cubic relations, together with non-terminating higher syzygies.

\subsection{Multiplicativity}
Of course, nothing prevents us from going in direction converse to the above.
We can compute the Hilbert series for the geometry of a VMS of a gauge theory whose syzygies encode the mesonic BPS spectrum and then attempt to find another gauge theory whose zeta function has the same enumeration.
In other words, we start from the upper left corner of the diagram in \eqref{diag}, trace to the right via plethystics, and then proceed contrariwise to the arrows in the bottom row, via localization to primes, and attempt to arrive at the VMS of another gauge theory.
In principle, we can proceed thence, forming another Hilbert series and another VMS, potentially {\it ad nauseam}.

This latter direction of reconstructing the zeta function, at least computationally, is more difficult than the one reconstructing the Hilbert series, because whereas plethystic exponentiation has an analytic inverse in terms of the M\"obius function, finding the factors in an Euler product is not so immediate.

First, one may ask why the above examples of elliptic curves and affine spaces  worked so nicely. A key is the multiplicativity. 
In order that a Dirichlet expansion be allowable in being developed into an Euler product, as was hinted in \S\ref{Lseries}, its coefficients $c_n$ in the series $\sum\limits_{n=1}^\infty \frac{c_n}{n^s}$ must be multiplicative; that is,
\begin{equation}\label{mult}
c_{mn} = c_m c_n \ ,
\end{equation}
 whenever $m$ and $n$ are coprime.
This can be seen by explicitly writing out the Euler product.
Note that this is a weaker condition from complete multiplicativity where this relation holds for all positive integers $m$ and $n$ (cf.~\cite{Hanany:2010cx} for multiplicativity in the context of enumerating D-brane orbifold theories).

In the case of affine space, the coefficients $c_n = n^k$ for some non-negative integer power $k$, which is certainly completely multiplicative and whence we were able to expand $\sum\limits_{n=1}^\infty \frac{n^k}{n^s} = \sum\limits_{n=1}^\infty \frac{1}{n^{s-k}} = \zeta(k-s)$. 
Subsequently, this allows for the Euler product over primes as $\prod\limits_{n=1} (1 - \frac{1}{p^{k-s}})^{-1}$, so that each of the factors is a rational function which can be then be interpreted as the zeta function of an algebraic variety.

Of course, in order that $c_n$ also be conceivable as the dimensions of the graded pieces in a commutative ring in accord with the Hilbert series and in addition to it being multiplicative, $c_n$ can only be such a pure monomial power. In other words, for example, the Dirichlet series of $\zeta(s) + \zeta(q-s)$ for some integer $q$ can be performed without difficulty, even though the coefficients $c_n = 1 + n^q$ will not be multiplicative, whereby prohibiting an Euler product with rational factors, and seems not amenable to an immediate arithmetic perspective.

This rather severe restriction, that the coefficients be both polynomial, in accord with Hilbert, and multiplicative, in accord with Hasse-Weil-Dirichlet, should not discourage us.
After all, for any gauge theory, especially those arising form branes at Calabi-Yau singularities, the entropy (asymptotic growth rate) of the BPS operators is entirely determined by the leading behaviour of the Hilbert polynomial which governs the Hilbert series -  a point to which we will return in the section on asymptotics.

\subsubsection{The Elliptic Curve}\label{s:E}
Nevertheless, examples still abound and let us continue with the train of thought prescribed above.
We have found, in \eqref{T2}, that the zeta function for $\IC$, or the affine cone over a single point, gave rise to a mysterious elliptic curve (the cone over which is the VMS of a gauge theory with two fields and constraints).
Thus, our starting point is the sextic curve at the upper left corner, giving us $a_n = n$.
We computed the plethystics in \eqref{macmahon}, but now let us compute the zeta function instead.

The arithmetic of an elliptic curve $E$ is a vast subject.
Luckily, we only require some rudiments.
First, from \eqref{Zp}, the local zeta function is of simple rational form:
\begin{equation}\label{Zelliptic}
Z_p(t; E) = \frac{1 - 2 a_p t + p t^2}{(1-pt)(1-t)} \ ,
\end{equation}
with a single parametre $a_p$ depending on the complex structure of the specific curve and on the prime $p$.
The global zeta function is therefore
\begin{equation}
Z(s; E) = \prod\limits_p Z_p(t = p^{-s}; E) = \zeta(s) \zeta(s-1) 
L(s; E)^{-1} \ .
\end{equation}
In forming this product we run into the issue of so-called good and bad reduction, as well as the concept of the {\bf Hasse-Weil L-Function}, formed by a product over primes dividing the {\bf conductor} of $E$. 
For a more detailed discussion we leave the reader to Appendix \ref{A:reduction}.

Our elliptic curve is a sextic and using $(x,y,z)$ as the weighted projective c\"oordinates of $W\IP^2_{[1:2:3]}$ in the given order, let us, for now, fix it to be $x^6 + y^3 + z^2 = 0$.
The L-function here is simply $L(s;E) =  \prod\limits_{p} (1 - 2a_p p^{-s} + p^{1-2s})^{-1}$.
The coefficient $a_p$ is determined only by $p$, {\it i.e.}, knowing the number of points of $E$ for $p^{r=1}$ determines the number for all $p^r$.
On equating \eqref{Zelliptic} with the definition \eqref{zeta} of the local zeta function gives such a relation (cf.~\cite{koblitz}):
\begin{equation}\label{Npa-rel}
N_{p^r} = p^r + 1 - \alpha^r - (p/\alpha)^r \ , 
\end{equation}
where $\alpha$ is the root for the numerator: $1 - 2 a_p t + p t^2 = (1-\alpha t)(1 - p/\alpha \ t)$.
In particular, $N_{p^1} = p + 1 - 2 a_p$, giving us, upon explicit enumeration of points, these following beginning values for $-a_p$:
\begin{equation}
\{0, 2, 9, 8, 54, 65, 135, 242, 252, 405, 404, 845, 819, 1070, 1080,
1377, 1710, 1409, 1682, 2484 \ldots \}
\end{equation}
Subsequently, we form the product over the local zeta function and apply the Dirichlet expansion, the first terms are:
\begin{equation}
Z(s; E) = 1 + \frac{3}{2^s} + \frac{8}{3^s} + \frac{9}{4^s} + \frac{24}{5^s}
+ \frac{24}{6^s} + \frac{24}{7^s} + \frac{21}{8^s} + \frac{32}{9^s}
+ \frac{72}{10^s} + \ldots
\end{equation}


\subsection{Modularity}
Having recoursed to elliptic curves, one could not possibly resist the opportunity to digress to modular forms.
The celebrated {\bf Taniyama-Shimura-Weil} Conjecture, now known as the Modularity Theorem by the works of Wiles et al., can be stated in explicit analytic form within our context. Let the L-series of an elliptic curve over $\IQ$, developed into a Dirichlet expansion $L(s) = \sum\limits_{n=1}^\infty \frac{a_n}{n^s}$, be recast into a generating function $\ell(q = e^{2 \pi i z}) = \sum\limits_{n=1}^\infty a_n q^n$, then this function is in fact a cusp form of weight 2 and level $N$, which is the conductor for the elliptic curve. We shall leave some more details explaining this correspondence to Appendix \ref{A:TS}.

Generalizing this modularity arising from the global zeta function for not just the Calabi-Yau one-fold, {\it viz.}, the elliptic curve, but to higher dimensions, has been a growing field \cite{modCY}. 
In a parallel spirit, that the mirror map, encoding the Gromov-Witten invariants for certain Calabi-Yau manifolds, especially non-compact toric Calabi-Yau threefolds exemplified in \S\ref{s:eg}, ({\it q.v.~e.g.}~\cite{Candelas:2000fq,Aganagic:2006wq}), should be a (quasi-) modular form, has also attained interest.

Indeed, our substitution, in accordance with the schematic diagram in the beginning of this section, of the L-series coefficients into the generating Hilbert series of the variety is very much in the spirit of this correspondence of Taniyama-Shimura et al.~and fall under the special situation when the growth rate of these coefficients be polynomial.
In fact, interpreting the plethystic exponential as a grand canonical partition function, one, as to which was earlier alluded, should call $t = e^w$, $\nu = e^{\mu}$ the {\bf fugacity} and $w$, $\mu$, the {\bf chemical potential} associated with the R-charge and number of colours of the SUSY gauge theory (q.v.~\cite{Feng:2007ur,Davey:2009et}).
In any saddle point analysis in extracting asymptotics, for example, as will be done in the ensuing section, contour integrals are to be performed with respect to these exponents.

In other words, in light of both modularity and the casting thereof as a fugacity when considering the theory as statistical-mechanical, the ``dummy variable'' $t$ in the Hilbert series of the vacuum variety of our gauge theory should be substituted exponentially, and a Fourier $q$-expansion, be afforded.
When the situation permits, then, the plethystic exponential of the Hilbert series, therefore becomes a modular form.

Of course, the Modularity Theorem is established only for elliptic curves and with some evidence compiling for higher dimension, so in our present context, it is expedient to re-consider our above example of the sextic elliptic curve in \S\ref{s:E}.
To facilitate the usage of \cite{magma}, let us de-homogenize $x^6 + y^3 + z^2 = 0$ and work with the 3 affine patches of the projective space $W\IP^2_{[1:2:3]}$.
First, in the patch $z=1$, we trivially have the quadric (defining $y' = y^3$) parabola $x^6 + y^3 + 1 = 0$ and the number of zeros over $\IF_{p}$ can be tabulated as $N_{p^{r=1}} = \{ 2, 3, 5, 3, 11, 9, 17, 39, 23, 29, 21, 45, 41, 57, 47, 53, 59, 45, 39, 71 \ldots \}$. Similarly, in the patch $y=1$, we have the cubic (defining $z' = z^2$) $x^6 + 1 + z^2 = 0$, and we have  $N_{p^{r=1}} = \{ 2, 4, 4, 0, 12, 8, 16, 36, 24, 28, 24, 56, 40, 60, 48, 52, 60, 32, 36,72 \ldots \}$.
In the patch $x=1$, however, we have the standard Weierstra\ss\ representation of the elliptic curve $1+y^3+z^2=0$ and that $N_{p^{r=1}} = \{2, 3, 5, 3, 11, 11, 17, 27, 23, 29, 27, 47, 41, 51, 47, 53, 59,\ldots\}$. Now, the conductor of the curve is found to be $144 = 2^4 \cdot 3^2$, thus the zeta function can be written as $Z(s) = \zeta(s) \zeta(s-1) \prod\limits_{p \nmid 144}
(1 - 2(N_p - p -1)p^{-s} + p^{1-2s})^{-1}$.

Correspondingly, using the Dirichlet coefficients of the L-function part of the above, and summing over the Fourier expansion, we find
\begin{eqnarray}
\nn
f(q) &=& 
q+4 q^7+2 q^{13}-8 q^{19}-5 q^{25}+4 q^{31}-10 q^{37}-8 q^{43}+\\
&&+
9 q^{49}+14 q^{61}+16q^{67}-10 q^{73}+4 q^{79}+8 q^{91}+14 q^{97}+\cO(q^{100})
\ ;
\end{eqnarray}
this is a cusp form of weight 2 and level 144, belonging to a vector space of dimension 59. Were these to be interpreted as not Fourier coefficients but, rather, power series coefficients encoding the non-terminating syzygies, we would be tempted to perform the plethystic exponentiation, formally with the variable $t$, and arrive at
\begin{eqnarray}
\nn
g(t) &=&
1+t+t^2+t^3+t^4+t^5+t^6+5 t^7+5 t^8+5 t^9+5 t^{10}+5 t^{11}+5 t^{12}+7t^{13}+
\\
&&+17 t^{14}+17 t^{15}+17 t^{16}+17 t^{17}+17 t^{18}+9 t^{19}+17
   t^{20}+\cO(t^{21})
\ .
\end{eqnarray}

As we saw in the case of the conifold, choice of complex structure of course sensitively affects the zeros. Experimenting with some values, we find that $z^2=y^3+16$ has the particularly small conductor of $27 = 3^3$, whereupon the associated L-series, and by Taniyama-Shimura-Wiles, the $q$-series of the associated cusp form, of weight 2 and level 27, is
{\hspace{-1cm}
\begin{eqnarray}
\nn
&&f(q) = q-2 q^4-q^7+5 q^{13}+4 q^{16}-7 q^{19}-5 q^{25}+2 q^{28}-4 q^{31}+11
   q^{37}+8 q^{43}-6 q^{49}-\\
&&~~~-10 q^{52}-q^{61}-8 q^{64}+5 q^{67}-7
   q^{73}+14 q^{76}+17 q^{79}-5 q^{91}-19 q^{97}+\cO(q^{100}) \ .
\end{eqnarray}
}
Luckily, the space of cusp forms of $\Gamma_0(27)$ at weight 2 is of dimension 1 and so is spanned by a single function \cite{finch}, thus we can actually write down the analytic form for $f(q)$, in terms of the Dedekind Eta function (the reciprocal of which, of course, without the prefactor, gives the standard partition of integers):
\begin{equation}
f(q) = \eta(q^3)^2 \eta(q^9)^2 \ , \qquad \mbox{with} \quad
\eta(q) := q^{\frac{1}{24}} \prod\limits_{n=1}^\infty (1 - q^n) \ .
\end{equation}

\section{Asymptot\ae \ Infinitorum}\label{s:asym}\setall
To the growth rate of the various coefficients relevant in our analyses we have alluded a number of times in our proceeding discussions, on account of summability, rationality as well as multiplicativity.
In this section, we embark on the examination of the asymptotics of the series central to our exposition.

The partition of integers is encoded by the Eta function, or in our language, by the Hilbert series for $\IC$ and the plethystics for the single-field free theory. The asymptotic behaviour, {\it i.e.}, the growth rate of the number of partitions for large integers, was determined by the celebrated result of Hardy and Ramanujan.
The generalization of this problem for weighted partitions was solved by Meinardus \cite{meinardus} and states that for the expansion
\begin{equation}
g(t) := \sum\limits_{n=0}^\infty d_n t^n
= PE[f(t)] = \prod\limits_{n=1}^\infty (1-t^n)^{-a_n} \ ,
\qquad \mbox{ with } \qquad
f(t) = \sum\limits_{n=0}^\infty a_n t^n \ ,
\end{equation}
the asymptotic behaviour of $d_n$ is:
\begin{equation}\label{mein}
d_n \sim C_1 n^{C_2} \exp
\left[ n^{\frac{\alpha}{\alpha+1}}(1+\frac{1}{\alpha})\left(A
\Gamma(\alpha+1) \zeta(\alpha+1)\right)^{\frac{1}{\alpha+1}}
\right] (1 + \cO(n^{-C_3})) \ .
\end{equation}
The constants, or critical exponents, in the above expression are determined as follows.
If the Dirichlet series for the coefficients $a_n$ of $f$,
{\it viz}., $D(s) := \sum\limits_{n=1}^\infty \frac{a_n}{n^s}$ with
${\rm Re}(s) > \alpha > 0$,
converges and is analytically continuable into the strip $-C_0 <
\mbox{Re}(s) \le \alpha$ for some real constant $0 < C_0 < 1$ and such
that in this strip, $D(s)$ has only 1 simple pole at $s = \alpha \in
\IR_+$ with residue $A$. The constants in \eqref{mein} are, explicitly,
\begin{eqnarray}
\nn
C_1 &=& e^{D'(0)} \frac{1}{\sqrt{2\pi(\alpha+1)}} \left(
    A \Gamma(\alpha+1)
    \zeta(\alpha+1)\right)^{\frac{1-2D(0)}{2(\alpha+1)}}, \\
C_2 &=& \frac{D(0)-1-\frac{\alpha}{2}}{\alpha+1} \ ,
\end{eqnarray}
and $C_3$ some positive constant with which we here need not contend.

Indeed, as mentioned in \eqref{mult}, the expansion coefficients $a_n$ of the Hilbert series grows polynomially and so asymptotically is governed by the leading term, which is essentially $a_n \sim K n^d$ where $d+1$ is the dimension of the VMS and $K$ is a constant coarsely depending on the geometry. For example, for affine space, $K=1$, and for orbifolds thereof, $K$ is some fraction depending on the order of the group.
We are thus led back to a situation very much akin to our example in \S\ref{s:cn}.
In this case, we have that $D(s) = K \zeta(s-d)$, $\alpha = d+1$, $A=K$, and
\begin{equation}
d_n \sim \frac{e^{K\zeta'(-d)}}{\sqrt{2\pi(d+2)}} 
   ((d+1)!\zeta(d+2))^{\frac{1-2K\zeta(-d)}{2(d+2)}} 
  n^{\frac{2K\zeta(-d)-d-3)}{2(d+2)}}\exp\left[K
  n^{\frac{d+1}{d+2}}\frac{d+2}{d+1} ((d+1)!\zeta(d+2))^{\frac{1}{d+2}}
  \right]  \ .
\end{equation}
Though standard to the Plethystic Programme, the above results can now be re-examined under our new light.
Indeed, \eqref{mein} dictates that the asymptotica of the total spectrum of (mesonic) BPS operators of a gauge theory is governed by the analytic characteristics - {\it viz.}, the placement of the pole and the residue thereon - of the Dirichlet series constructed from the coefficients of the Hilbert series of its VMS.
However, this is precisely the duality substitution outlined in diagram \eqref{diag} and on which we dwelled in \S\ref{s:compare}.

Carrying on with this train of thought, we should ascribe a geometry to $D(s)$;
this, of course, is one for which the global zeta function is $Z(s) = K \zeta(s-d)$.
The local zeta function is thus $Z_p(t) = K(1-p^dt)^{-1}$, which, normalizing the $\log K$ constant term, gives $\IC^d$. Therefore, recalling the origin of our coefficients $a_n$, asymptotically then, we have that this $d$-dimensional gauge theory arising from arithmetic is dual ``holographically'' to the $d+1$ dimensional gauge theory emerging from geometry.

Having entered the vast realm of analytic and asymptotic properties of zeta functions, one could hardly contain oneself, as a parting speculation, from remarking on the r\^ole of its zeros and poles.
The celebrated zeros of the Riemann zeta function aside, it being the analytic continuation of the Hasse-Weil zeta function for a single point, more contiguous to our present theme is perhaps the Conjecture of {\bf Birch and Swinnerton-Dyer} (cf. \cite{koblitz}).
It states that for an elliptic curve (and possibly generalizing to higher Abelian varieties) of rank\footnote{According to Mordell-Weil, the points $E(\IQ)$ of an elliptic curve $E$ over $\IQ$ form a group which decomposes as $E(\IQ) \simeq E(\IQ)_{tor} \oplus \IZ^r$ where $E(\IQ)_{tor}$ is the torsion part, constituted by some finite group and $r$, called the rank, governs the number of copies of $\IZ$. Hence $r>1$ means that there are, in particular, infinite rational points on $E$.
} $r$, the L-function tends as $L(s;E) \sim c (s-1)^r$ for some constant $c$ and near $s=1$. Thus, if the L-function vanishes at 1, then there is an infinite number of rational points.

Now, we have argued above that the asymptotic growth rate of the BPS spectrum of a gauge theory is controlled by the pole and residue of the Dirichlet series arising from a ``holographic'' dual, or equivalently, by the order of the zero of the reciprocal of the Dirichlet series.
The reciprocal of the Hasse-Weil zeta function clearly inverts each local zeta factor. 
This, as was pointed out in \cite{Candelas:2004sk}, is an interesting action: it exchanges even and odd cohomology according to the Weil Conjectures.
For Calabi-Yau spaces, or any generalization thereof, this is actually {\bf mirror symmetry}.
Eq (10.6) in {\it ibid.}~proposes a ``quantum zeta function'' $Z_p(t; \cM)^Q$ for a Calabi-Yau manifold $\cM$ whose mirror is $\cW$ such that the numerator of $Z_p(t; \cM)^Q$ is that of the numerator of the ordinary $Z_p(t; \cM)$ and the denominator is the numerator of $Z_p(t; \cW)$. 
Therefore, since the delocalization to the global zeta function is via a product, the zero of the Hasse-Weil zeta function of one manifold is the pole of that of its mirror.

Now, the order of the zero at 1, at least for Abelian varieties, determines the rank of its group of rational points by (generalizations of) Birch-Swinnerton-Dyer. 
This then, should in turn be ascertained by the order of the pole at 1 for the mirror variety.
However, by our correspondences in the preceding discussions in \S\ref{s:compare}, the pole of the Dirichlet series representation of the global zeta function determines the asymptotic growth, in the manner of a critical exponent, of a gauge theory whose VMS possesses a Hilbert series which can be identified formally with the coefficients in the Dirichlet expansion.
In this fashion, the asymptotics of one gauge theory, whose VMS is $\cM$, via the pole structure of the plethystic exponential of the fundamental generating function of its BPS spectrum, would control the zeroes of the zeta function of another theory whose VMS is the mirror $\cW$ of $\cM$, and thence, the rank of the rational points on $\cW$.
It is of course interesting to pursue this line of thought, however, for now, let us content ourselves with leaving this to future work.

\section{Prospectus}\label{s:conc}
We have embarked on a somewhat length journey through the expansive landscape of supersymmetric gauge theories, treading along a selected path which is guided by two principles, the first geometric, hinging upon the established Plethystic Programme for the counting of the BPS spectrum of the operators, and the second arithmetic, founded on the zeros of the vacuum moduli space over number fields of finite characteristic.
Drawing from the observations over a plenitude of examples, we have attempted to regard the two parallel enumerative problems under the same light, as outlined by the diagram in \eqref{diag}.
Both proceed from an intrinsic property, exponentiated to arrive at a fundamental rational generating function, and then exponentiated again in order to be delocalized to an infinite product: the former, originates from the syzygies and ends with a canonical partition function and the latter, begins with the zeroes over finite fields and arrives at an Euler product over primes.

By explicitly constructing pairs of gauge theories, where the vacuum of one governs the other by having their generating functions exchanged, we have observed an interesting duality wherein the geometry of one and the arithmetic of another inter-mingle.
Asymptotic analyses on the growth rates of the coefficients in the generating functions, by construction constituting the enumerations, suggest a curiously ``holographic'' nature of this duality, which holds for arbitrary gauge theories, as coursely controlled asymptotically by the dimension of the vacuum moduli space.
In due course, we have been inevitably led to the study of L-functions, Dirichlet series, analytic behaviour of the zeroes and poles of Hasse-Weil zeta functions, as well as the Modularity Theorem, all of which tremendous subjects in themselves; we pray that the patient reader has as much forgiven our inexpertise as he or she has indulged in our long exposition, especially in the drudgery of our examples and experimentation.

It is hoped that we have only skimmed over the surface of a deeper subject.
To mention but a few prospects in this brief epilogue, we should, for instance, explore the full quantum moduli space, which in the geometry often materializes as deformations in complex structure.
We have seen how such deformations could drastically alter, if not the chiral ring, at least the arithmetic.
Moreover, the vacuum moduli space of gauge theories, especially in the context of D-brane world-volume physics, should be comprehended scheme-theoretically, with the classical Abelian case being the centre of some non-commutative algebra.
We have placed rational restrictions on the Hilbert series and the local zeta functions because of commutative algebraic geometry, however, the realm of non-commutativity would significantly relax such constraints and would, naturally, lead to further correspondences.
For now, let us repose awhile from our excursions onto this territory of physics, geometry and number theory, and regain our strength by further reflections, before voyaging further on such fertile ground.

\section*{Acknowledgements}
\begin{spacing}{1}
{\it {\small
Pro commentariis amicorum carissimorum suorum, Philippi Candelas, Xeniae de la Ossa, Amihayi Hanany, Vishnuus Jejjala, Iacobi Moxness et Georgii Minic, Y.-H.H.~gratias plurimas agit, dum debet animam suam Scientiae et Technologiae Concilio Anglicae, et Ricardo Fitzjames, Episcopo Londiniensis, ceterisque omnibus benefactoribus Collegii Mertonensis Oxoniensis, Universitate Urbis Londiniensis, et Cathedrae Professoris Chang-Jiangiensis Universitatis Nan-Kai \footnote{
Work supported in part by the STFC, UK in association with the Rudolf Peierls Centre for Theoretical Physics, University of Oxford, a Supernumerary, quandam FitzJames, Fellowship of Merton College, Oxford, as well as an impending Readership from City University, London and a Chang-Jiang Chair Professorship from the Chinese Ministry of Education, at Nan-Kai University, Tian-Jin.
}.
\\[0.05in]
Bibet pro salutatem Dominae Iessicae Rawson, custodis Mertonensis, salutem dicetque Domino Martino Taylor, custodi novi. Stet Fortuna Domus!
\\[0.05in]
Sed super omnes, Domino illuminatione mea, pro amore Catharinae Sanctae Alexandriae, lacrimarum Mariae semper Virginis, et ad Maiorem Dei Gloriam hoc opusculum dedicat.
}}
\end{spacing}

\appendix
\section{Primes of Good and Bad Reduction}\label{A:reduction}
In performing the product over all the prime in obtaining the global zeta function from the local, we encounter situations of primes of bad-reduction where the variety may become singular.
We illustrate this, in a pedagogical manner, using an explicit example explained lucidly by \cite{koblitz} whose excellent presentation we will attempt to follow here.

Take the elliptic curve $\{y^2 = x^3 - n^2 x \} \subset \IC[x,y]$, with $n$ some integer parametre, and denote it as $E_n$.
Indeed, the curve, together with its Jacobian, prescribe the simultaneous system: $\{y^2 - x^3 + n^2 x = 0, 2y = 0, 3x^2-n^2 = 0\}$.
Working over a field of characteristic $p^r$ , this has non-trivial solutions if $p | 2n$ (for $p=2$, $(\pm 1,0)$ and for $p | n$, $(0,0)$), whereby making the point corresponding to the solution singular and the elliptic curve, degenerate. Such a prime $p$ is called one of {\bf bad reduction}.
Over these primes, there are always $N_{p^r} = p^r + 1$ points.
For instance, when $p |n$, the curve degenerates to a complex line $y^2 = x^3$, over which, we recall, there are $p^r + 1$ points over $\IF_{p^r}$.
Over the remaining primes, of good reduction, the Weil conjectures give us the rational form in \eqref{Zelliptic}.
Note that the bad reduction primes is always a finite set, determined as those divisible by some parametre. Here $2n$ is this governing parametre and gives rise to the so-called {\bf conductor} for the given elliptic curve. In general, the conductor $N$ is a single integer whose prime factors are precisely those of bad reduction. In summary,
\begin{equation}
Z_p(t; E_n) = \left\{
\begin{array}{rcl}
\frac{1 - 2 a_p t + p t^2}{(1-pt)(1-t)} & & p \nmid 2n \ , \\
\frac{1}{(1-pt)(1-t)} & & p \mid \ 2n \ ,
\end{array}
\right.
\end{equation}
so that the global zeta function, upon substituting $t = p^{-s}$, should be
\begin{equation}
Z(s; E_n) = 
\prod\limits_{p \nmid 2n} \frac{1 - 2a_p t + p t^2}{(1-pt)(1-t)}
\prod\limits_{p \mid \ 2n} \frac{1}{(1-pt)(1-t)}
=
\frac{\zeta(s)\zeta(s-1)}{L(s; E_n)} \ ,
\end{equation}
where $L(s; E_n) := \prod\limits_{p \nmid 2n} (1 - 2a_p p^{-s} + p^{1-2s})^{-1}$ is dubbed the {\bf Hasse-Weil L-Function}.
The relation between the parametre $2a_p$ and the number of points over $p^{r=1}$ is as in \eqref{Npa-rel}:
\begin{equation}\label{Npa-rel2}
2a_p(E_n) = p + 1 - N_{p} \ .
\end{equation}

\section{Modularity and Hasse-Weil}\label{A:TS}
The particular case of Taniyama-Shimura, now called the Modularity Theorem by the work of Wiles et al.~ special case of the Langlands Programme, which has of late become important in string theory due to Witten - upon which we here briefly touch is the Hasse-Weil Conjecture for elliptic curves.
Of course, all these impinge on an enormous subject, for which we have neither the qualifications nor the space to expound in any depth.
We shall concentrate on the analytics of the Dirichlet expansion  $L(s; E) = \sum\limits_{n=1}^\infty \frac{a_n}{n^s}$ for the L-function of a given elliptic curve $E$.
To set notation, by a modular of weight $k$ and level $N$ we mean an analytic function $f(z)$ defined on the complex upper-half plane $\{z \in \IC : \im(z) >0\}$ such that $f\left( \frac{az+b}{cz+d}\right) = (cz+d)^k f(z)$ under $\Gamma_0(N)$, a particular congruence subgroup of the modular group $SL_2(\IZ)$ where $c$ divides $N$.
Furthermore, a cusp form is one for which the zeroth coefficient $a_0$ of its $q$-expansion $f(z) = \sum\limits_{n=1}^\infty a_n q^n$ (where $q := e^{2\pi i z}$) vanishes.

Now, the $E_n$  in Appendix \ref{A:reduction} have rather complicated conductors, so let us use a simpler example.
Consider the elliptic curve $y^2 + x y + y = x^3 - x^2 - x$, one could find \cite{magma} that the conductor is 11.
We can tabulate, for some first primes, some leading values of $N_p$ (which we can readily find on the computer) and the coefficient $a_p$ in the local zeta function, related thereto by a relation in analogy to \eqref{Npa-rel2}: $a_p = p - N_p$:
\begin{equation}\begin{array}{c||ccccccccccccccc}
p & 2 & 3 & 5 & 7 & 11 & 13 & 17 & 19& 23& 29& 31& 37& 41& 43& 47 \\ \hline
N_p & 3& 3& 7& 3& 11& 15& 16& 23& 19& 23& 27& 39& 47& 39& 47 \\
a_p &-1& 0& -2& 4& 0& -2& 1& -4& 4& 6& 4& -2& -6& 4& 0 \\
\end{array}\end{equation}
Forming the L-function and expanding into Dirichlet series gives:
\begin{eqnarray}
\nn
L(s) & = & \prod\limits_{p \ne 11} (1 - a_p p^{-s} + p^{1-2s})^{-1} = 
\sum\limits_{n=1}^\infty \frac{c_n}{n^s} \ ; \\
& & c_n = \{1, -1, 0, -1, -2, 0, 4, 3, -3, 2, 0, 0, -2, -4 \ldots \}
\end{eqnarray}
On the other hand, the space of modular forms at each weight is a finitely generated vector space. In particular, cusp forms of weight 2 and level 17 is generated by a single function, given as a plethystic type product, with $\eta(q) := q^{\frac{1}{24}} \prod\limits_{n=1}^\infty (1 - q^n)$ the standard Dedekind eta-function:
\begin{eqnarray}
\nn
\frac{\eta(q)\eta(q^4)^2\eta(q^{34})^5}{\eta(q^2)\eta(q^{17})\eta(q^{68})^2}
-
\frac{\eta(q^2)^5\eta(q^{17})\eta(q^{68})^2}{\eta(q)\eta(q^{4})^2\eta(q^{34})}
 =
\sum_{n=1}^\infty a_n q^n \ ; && \\
a_n = \{1, -1, 0, -1, -2, 0, 4, 3, -3, 2, 0, 0, -2, -4  \ldots\} && \ . 
\end{eqnarray}
As one can see, the coincidence of the two sets of expansion coefficients is highly non-trivial and constitutes one of the highlights of twentieth century mathematics.


\end{document}